\documentclass[useAMS,usenatbib]{mn2e}
\usepackage{graphicx}
\usepackage{epsfig}
\usepackage{epsf}
\usepackage{amsmath}
\usepackage{amssymb}
\usepackage{stfloats}
\usepackage{aas_macros}
\usepackage{deluxetable}
\pdfoutput=0
\newcommand\ion[2]{#1$\,${\small{#2}}\relax}
\title[Twin SNe\,Ia 2011fe and 2011by]{Twins for life? A comparative analysis of the Type Ia supernovae 2011fe and 2011by}

\author[Graham et al.]{M.~L. Graham$^{1}$\thanks{E-mail: melissagraham@berkeley.edu}, R.~J. Foley$^{2,3}$, W. Zheng$^{1}$, P.~L. Kelly$^{1}$, I. Shivvers$^{1}$, \newauthor J.~M. Silverman$^{4}$, A.~V. Filippenko$^{1}$, K.~I. Clubb$^{1}$, and M. Ganeshalingam$^{5}$ \\
$^{1}$ Department of Astronomy, University of California, Berkeley, CA 94720-3411 USA \\
$^{2}$ Astronomy Department, University of Illinois at Urbana-Champaign, 1002 W.\ Green Street, Urbana, IL 61801 USA \\
$^{3}$ Department of Physics, University of Illinois at Urbana-Champaign, 1110 W.\ Green Street, Urbana, IL 61801 USA \\
$^{4}$ Department of Astronomy, University of Texas at Austin,
Austin, TX 78712 USA \\
$^{5}$ Lawrence Berkeley National Laboratory, 1 Cyclotron Road, MS 90R4000, Berkeley, CA 94720 USA \\
}

\begin{document}
\pagerange{\pageref{firstpage}--\pageref{lastpage}} \pubyear{2014}

\maketitle

\label{firstpage}

\begin{abstract}
The nearby Type Ia supernovae (SNe\,Ia) 2011fe and 2011by had nearly identical photospheric phase optical spectra, light-curve widths, and photometric colours, but at peak brightness SN\,2011by reached a fainter absolute magnitude in all optical bands and exhibited lower flux in the near-ultraviolet (NUV). Based on those data, Foley \& Kirshner (2013) argue that the progenitors of SNe 2011by and 2011fe were supersolar and subsolar, respectively, and that SN\,2011fe generated 1.7 times the amount of $^{56}$Ni as SN\,2011by. With this work, we extend the comparison of these SNe\,Ia to 10 days before and 300 days after maximum brightness with new spectra and photometry. We show that the nebular phase spectra of SNe\,2011fe and 2011by are almost identical, and do not support a factor of 1.7 difference in $^{56}$Ni mass. Instead, we find it plausible that the Tully-Fisher distance for SN\,2011by is an underestimate, in which case these SNe\,Ia may have reached similar peak luminosity, formed similar amounts of $^{56}$Ni, and had lower metallicity progenitors than previously estimated. Regardless of the true distance to SN\,2011by, we find that the {\it relative} progenitor metallicity difference remains well supported by their disparity in NUV flux, which we show to be even stronger at pre-maximum epochs --- although contributions from differences in total ejecta mass, viewing angle, or progenitor density cannot be ruled out. We also demonstrate that, independent of distance modulus, SN\,2011by exhibits a late-time luminosity excess that cannot be explained by a light echo, but is more likely to be the result of greater energy trapping by the nucleosynthetic products of SN\,2011by.
\end{abstract}


\section{Introduction} \label{sec:intro}

Type Ia supernovae (SNe\,Ia), characterised by the lack of hydrogen and helium and the presence of silicon in their optical spectra (e.g., Filippenko 1997), are standardisable candles through the empirical width-luminosity relation (WLR; Phillips 1993). This intrinsic diversity in peak luminosity is attributed to variations in the amount of radioactive nickel ($^{56}$Ni) synthesised in the explosion (Arnett 1982). After the WLR, the empirical calibration incorporates a second term based on SN colour. This is attributed mainly to host-galaxy dust extinction, but also includes intrinsic colour variations among SNe\,Ia. In order to standardise SNe\,Ia as fully as possible, it is necessary to understand the physical causes of the observed variations, especially those that may evolve over cosmological timescales. In particular, there may be a relationship between peak luminosity and progenitor metallicity, where the additional neutrons from a higher abundance of heavy elements could lead to more stable nuclear burning and relatively less radioactive $^{56}$Ni. 

For this reason, the effect of progenitor metallicity on the photometric and spectroscopic evolution of SNe\,Ia is an important and active topic of research. Theoretical and analytical efforts to understand it include the models of Lentz et al. (2000), who show that higher metallicity produces a depressed flux in the near-ultraviolet (NUV) without affecting the optical spectrum, and Sauer et al. (2008), who find that higher metallicities can instead enhance the NUV flux by scattering photons to bluer wavelengths. Observational efforts have shown that the NUV diversity of SNe\,Ia cannot be attributed solely to dust and is likely related to progenitor metallicity (Ellis et al. 2008), and that progenitor metallicity could account for $\sim 10\%$ of the variation in $^{56}$Ni mass in SNe\,Ia (Howell et al. 2009). Comparisons of low-redshift ($z \approx 0$) to intermediate-redshift ($z\approx 0.2$--0.6) SN\,Ia samples find that low-$z$ SNe\,Ia have similar optical spectra but lower NUV flux, and that the NUV flux depression agrees with model predictions for a progenitor metallicity difference consistent with universal chemical enrichment between sample redshifts (Foley et al. 2012; Maguire et al. 2012; Walker et al. 2012). In general, theory and observations both appear to agree that progenitor metallicity is correlated with the NUV flux of SNe\,Ia.

As a case study for the effects of progenitor metallicity on SNe\,Ia, Foley \& Kirshner (2013; hereafter FK13) present a comparative analysis of two nearby SNe\,Ia: SNe\,2011fe and 2011by. They demonstrate how the near-maximum spectra are nearly identical in their optical features, but that SN\,2011by exhibits a significantly depressed NUV flux compared to SN\,2011fe. These SNe\,Ia are also very similar in optical light-curve width and colour, but SN\,2011by appears intrinsically fainter by $\sim0.6$\,mag, corresponding to a ratio of synthesised $^{56}$Ni mass of $M_{\rm fe}/M_{\rm by} = 1.7^{+0.7}_{-0.5}$. FK13 show that the disparity in NUV flux and peak luminosity cannot be explained by dust absorption in the host galaxy of SN\,2011by, because this would also redden the optical spectrum. They instead argue that the differences in NUV flux and peak luminosity together indicate that SNe\,2011fe and 2011by had progenitor stars below and above solar metallicity, respectively, and claim the first robust detection of different metallicities for SN\,Ia progenitors.

In this work we integrate new and previously published photometry and spectroscopy for SNe\,2011fe and 2011by, covering phases $-10$ to +300 days relative to peak brightness, that lead to significant new insights about the physical nature of these two twin-like SNe\,Ia. We present the observations in \S \ref{sec:obs} and include a brief assessment of the host-galaxy properties in \S \ref{ssec:host}. Our analysis of the data is described in \S \ref{sec:ana}, including the photometry (\ref{ssec:anaphot}), the photospheric phase spectra (\ref{ssec:anaspec}), and the nebular phase spectra (\ref{ssec:ananeb}). We discuss the progenitor metallicity explanation for the NUV flux disparity in \S \ref{ssec:discphot}, and demonstrate in \S \ref{ssec:discneb} how the nebular spectra are evidence for an underestimated Tully-Fisher distance modulus for SN\,2011by. A comprehensive discussion of alternative models and explanations for the NUV flux diversity is provided in \S \ref{ssec:discaltpm} and \S \ref{ssec:discalt}, and we discuss the physical implications for the late-time luminosity excess of SN\,2011by in \S \ref{ssec:discLC}. Section \ref{sec:con} summarises our results and conclusions.

\begin{table}\begin{center}\begin{minipage}[bp]{3.2in} \setlength{\tabcolsep}{2.pt}
      \caption{Early-Time SN Parameters. \label{tab:params}}
\centering
  \begin{tabular}[bp]{@{}lcc@{}} 
  \hline
  \hline
Parameter & SN\,2011fe & SN\,2011by \\
\hline
Peak Date [MJD]               & $55815.5\pm0.3$$\rm^{R}$         & $55690.6\pm0.1$$\rm^{M}$   \\
$m_B(\rm{peak})$ [mag]        & $10.00\pm0.02$$\rm^{R}$          & $12.89\pm0.03$$\rm^{S}$    \\
$\Delta m_{15}(B)$ [mag]      & $1.12\pm0.05$$\rm^{R,V}$         & $1.14\pm0.03$$\rm^{S}$     \\
Rise Time [days]              & $18.8\pm0.3$$\rm^{R,N}$          & $18.2\pm0.3$                 \\
Distance Modulus [mag]        & $29.04\pm0.05$$\rm^{SS}$         & $31.34\pm0.36$$\rm^{T}$      \\
$M_B(\rm{peak})$ [mag]             & $-19.21\pm0.15$$\rm^{R}$         & $-18.45\pm0.39$ \\
$M_{\rm{Ni}^{56}}$ [M$_{\odot}$]  & $0.53\pm0.11$$\rm^{P2}$          & $\sim$$0.31^{+0.19}_{-0.16}$$\rm^{F}$ \\
Photospheric Velocity         & normal$\rm^{P}$                  & normal$\rm^{S}$ \\
Velocity Gradient Group       & LVG$\rm^{P}$                     & LVG$\rm^{S}$ \\
\hline
\end{tabular}
\vspace{-24pt}
\tablenotetext{}{F=Foley et al. 2013; N=Nugent et al. 2011; M=Maguire et al. 2012; P=Parrent et al. 2012; P2=Pereira et al. 2013; R=Richmond \& Smith 2012; SS=Shappee \& Stanek 2011; S=Silverman et al. 2013; T=Tully et al. 2009; V=Vinko et al. 2012}
\end{minipage} \end{center}
\end{table}

\begin{table*}
\begin{center}
\begin{minipage}[bp]{5.2in}
\setlength{\tabcolsep}{2.pt}
\caption{List of Spectroscopy Used in this Work \label{tab:spec}}
\centering
  \begin{tabular}[bp]{@{}lccccc@{}} 
  \hline
  \hline
SN Name & Date & SN\,Ia Phase & Telescope & Gratings & Exposure Time  \\
        &  &  [days from peak] & and Instrument &  and Grisms      & [s] \\ 
\hline
SN\,2011by & 2011-04-28 & -12  & Shane+Kast & 600/4310,300/7500 & 600       \\
           & 2011-04-30 & -10  & HST+STIS   & G230L,G430L       & 2332,2263 \\
           & 2011-05-04 &  -6  & Shane+Kast & 600/4310,300/7500 & 600       \\
           & 2011-05-05 &  -5  & HST+STIS   & G430L             & 2263      \\
           & 2011-05-09 &  -1  & HST+STIS   & G230L,G430L       & 2332,2263 \\
           & 2011-05-10 &   1  & Shane+Kast & 600/4310,300/7500 & 600       \\
           & 2011-05-18 &   9  & HST+STIS   & G430L             & 2263      \\
           & 2011-12-02 &  206 & Keck+LRIS  & 600/4000,400/8500 & 450       \\
           & 2012-03-15 &  310 & Keck+LRIS  & 600/4000,400/8500 & 900       \\
\hline
SN\,2011fe & 2011-08-31 & -12 & HST+STIS   & G230L,G430L,G750L  & 1100,1156,820  \\
           & 2011-09-03 &  -9 & HST+STIS   & G230LB,G430L,G750L & 1320,195,195   \\
           & 2011-09-07 &  -5 & HST+STIS   & G230LB,G430L,G750L & 1060,160,160   \\
           & 2011-09-10 &  -2 & HST+STIS   & G230LB,G430L,G750L & 1060,160,160   \\
           & 2011-09-13 &   2 & HST+STIS   & G230LB,G430L,G750L & 830,160,140    \\
           & 2011-09-19 &   8 & HST+STIS   & G230LB,G430L,G750L & 1000,160,140   \\
           & 2012-04-23 &  224 & Shane+Kast & 600/4310,300/7500  & 450            \\
           & 2012-07-17 &  309 & Shane+Kast  & 600/4310,300/7500  & 3000          \\
\hline
\end{tabular}
\end{minipage} \end{center}
\end{table*}

\section{Observations} \label{sec:obs}

Here we present the new photometry and spectroscopy that we incorporate with published observations of SNe\,2011fe and 2011by and their host galaxies.

\subsection{SNe\,2011fe and 2011by}

SN\,2011fe was discovered by the Palomar Transient Factory (Rau et al. 2009; Law et al. 2009; internal designation PTF11kly) on 24 Aug. 2011 (UT dates are used throughout this paper) at $\alpha$(J2000) = $14^{\rm h} 03^{\rm m} 05.81^{\rm s}$, $\delta$(J2000) = $+54^\circ 16' 25.4''$. SN\,2011fe was classified as a young, spectroscopically normal SN\,Ia (Nugent et al. 2011). Observations of SN\,2011fe revealed that the total reddening from both the Milky Way and its host, M101, was $E(B-V)\lesssim0.05$\,mag, and that the circumstellar environment of SN\,2011fe was ``clean" (Chomiuk et al. 2012; Horesh et al. 2012; Patat et al. 2013; Pereira et al. 2013). For this work we use only the line-of-sight Galactic extinction values $A_B=0.031$, $A_V=0.023$, $A_R=0.019$, and $A_I=0.013$\,mag from the NASA Extragalactic Database (NED\footnote{http://ned.ipac.caltech.edu/}; Schlafly \& Finkbeiner 2011; Schlegel, Finkbeiner \& Davis 1998). 

The photometric evolution of SN\,2011fe was presented by Richmond \& Smith (2012); although they report a decline parameter $\Delta m_{15} (B) = 1.21\pm0.03$\,mag, Vinko et al. (2012) find that this is probably a misprint because they fit the same data and find $\Delta m_{15} (B) = 1.12\pm0.05$\,mag. Richmond \& Smith (2012) also report a peak apparent magnitude of $m_B = 10.00\pm0.02$ on 11.5 Sep. 2011, and a peak absolute magnitude of $M_B = -19.21\pm0.15$. By $\sim30$ days after maximum brightness, SN\,2011fe was at very high airmass (4--5) and the photometric calibration deteriorates (especially in the $B$ band), but this does not affect the measurements of peak magnitude or $\Delta m_{15}(B)$. The early optical spectroscopic evolution of SN\,2011fe was presented by Parrent et al. (2012), who classify it as spectroscopically normal and a member of the ``low velocity gradient" group, referring to the change in velocity of the \ion{Si}{II} $\lambda$6355 line (Benetti et al. 2005).

SN\,2011by was discovered by the Xingming Observation Sky Survey on 26 April 2011 (Jin \& Gao 2011), at $\alpha$(J2000) = $11^{\rm h}55^{\rm m}45.56^{\rm s}$, $\delta$(J2000) = $+55^\circ 19' 33.8''$. It was spectroscopically classified as a young SN\,Ia by Zhang et al. (2011). For this work we use the line-of-sight Galactic extinction values $A_B=0.051$, $A_V=0.038$, $A_R=0.030$, $A_I=0.021$\,mag (NED; Schlafly \& Finkbeiner 2011; Schlegel, Finkbeiner \& Davis 1998). The photometric and spectroscopic evolution of SN\,2011by was presented by Silverman et al. (2013), who determined a decline parameter $\Delta m_{15} (B) = 1.14\pm0.03$\,mag and show that it is spectroscopically normal and also a member of the ``low velocity gradient" class. For this work we use the peak $B$-band magnitude date from Maguire et al. (2012), as did FK13, of 9.6 May 2011.

Interestingly, Milne et al. (2013) present a study of the NUV-optical colours of a sample of 23 SNe\,Ia, finding that both SNe\,2011fe and 2011by fall in the ``NUV-blue" category. All SNe\,Ia in this category exhibit a ``low velocity gradient" in their \ion{Si}{II} $\lambda$6355 line, and they exhibit carbon lines at early times. We show in later sections that SNe\,2011fe and 2011by are consistent in these respects as well. The main characteristics of SNe\,2011fe and 2011by are listed in Table \ref{tab:params}.

\subsection{Host-Galaxy Properties}\label{ssec:host}

The host galaxy of SN\,2011fe is M101 (NGC 5457), for which a Cepheid distance of $6.4\pm0.7$\,Mpc ($\mu = 29.04\pm0.23$\,mag) was measured by Shappee \& Stanek (2011). The recession velocity of M101 is $v_{\rm rec}=241$ $\rm km\ s^{-1}$ (de Vaucouleurs et al. 1993). The host galaxy of SN\,2011by is NGC 3972, an inclined spiral, and the most recent distance measurement listed in NED is 18.5\,Mpc ($\mu = 31.34\pm0.36$\,mag) from the Tully-Fisher relation (Tully et al. 2009). The host-galaxy recession velocity is $v_{\rm rec}=852$ $\rm km\ s^{-1}$ (Verheijen \& Sancisi 2001).

The distance modulus of SN\,2011by derived from its light-curve fit, $\mu=32.01\pm0.07$\,mag (NED; Maguire et al. 2012), is $\sim0.7$\,mag higher than the Tully-Fisher distance modulus to its host, NGC 3972. Could the Tully-Fisher distance be an underestimate of the true distance? Although it is supported by multiple measurements from independent publications, the uncertainty is relatively large, and among these measurements the redder filters produce larger distance moduli; in fact, the highest Tully-Fisher distance modulus of $\mu=31.82\pm0.45$\,mag is measured from the $K$ band. Given that NGC 3972 is an edge-on spiral with obvious dust lanes, it is reasonable to suspect that the Tully-Fisher distance modulus may underestimate the distance to SN\,2011by. If so, SNe\,2011fe and 2011by may have reached similar peak absolute magnitudes, as suggested by their similar light-curve widths. Throughout this paper we discuss how this would change the physical interpretations of $^{56}$Ni mass and progenitor metallicity for these SNe\,Ia. To establish a more reliable distance to NGC 3972, we have an approved {\it Hubble Space Telescope (HST)\/} Cycle 22 programme to obtain a Cepheid distance measurement. 

As described by FK13, M101 and NGC 3972 are spiral galaxies with gas-phase metallicities of $\rm 12 + \log(O/H) = 9.12$ and 8.97, respectively (SDSS spectra, using a solar value of 8.86; Prieto et al. 2008). Stoll et al. (2011) apply a metallicity gradient to M101 and find that the metallicity at the position of SN\,2011e is subsolar, at $\rm 12 + \log(O/H) = 8.45$. We point out that SN\,2011by is also well offset from the centre of NGC 3972: 5.3\arcsec\ east and 19.1\arcsec\ north. Based on the semimajor and semiminor axes, if SN\,2011by is in the disc, it has a projected radial offset of half the effective radius of NGC 3972. Spiral galaxies commonly have a metallicity lower by $\Delta({\rm 12+\log(O/H)}) \approx -0.5$ at this radius (Zaritsky et al. 1994). This would put the local environment metallicity for SN\,2011by at $\rm 12 + \log(O/H) \approx 8.47$, in agreement with the estimate for SN\,2011fe. However, a difference of $\rm \Delta(\log(O/H)) \approx 0.5$, or $Z$/$\rm Z_{\odot} \approx 3$, in progenitor metallicity certainly cannot be excluded based on these estimates.

\begin{figure*}
\begin{center}
\includegraphics[trim=1.2cm 0cm 0.2cm 0cm,clip=true,width=8.5cm]{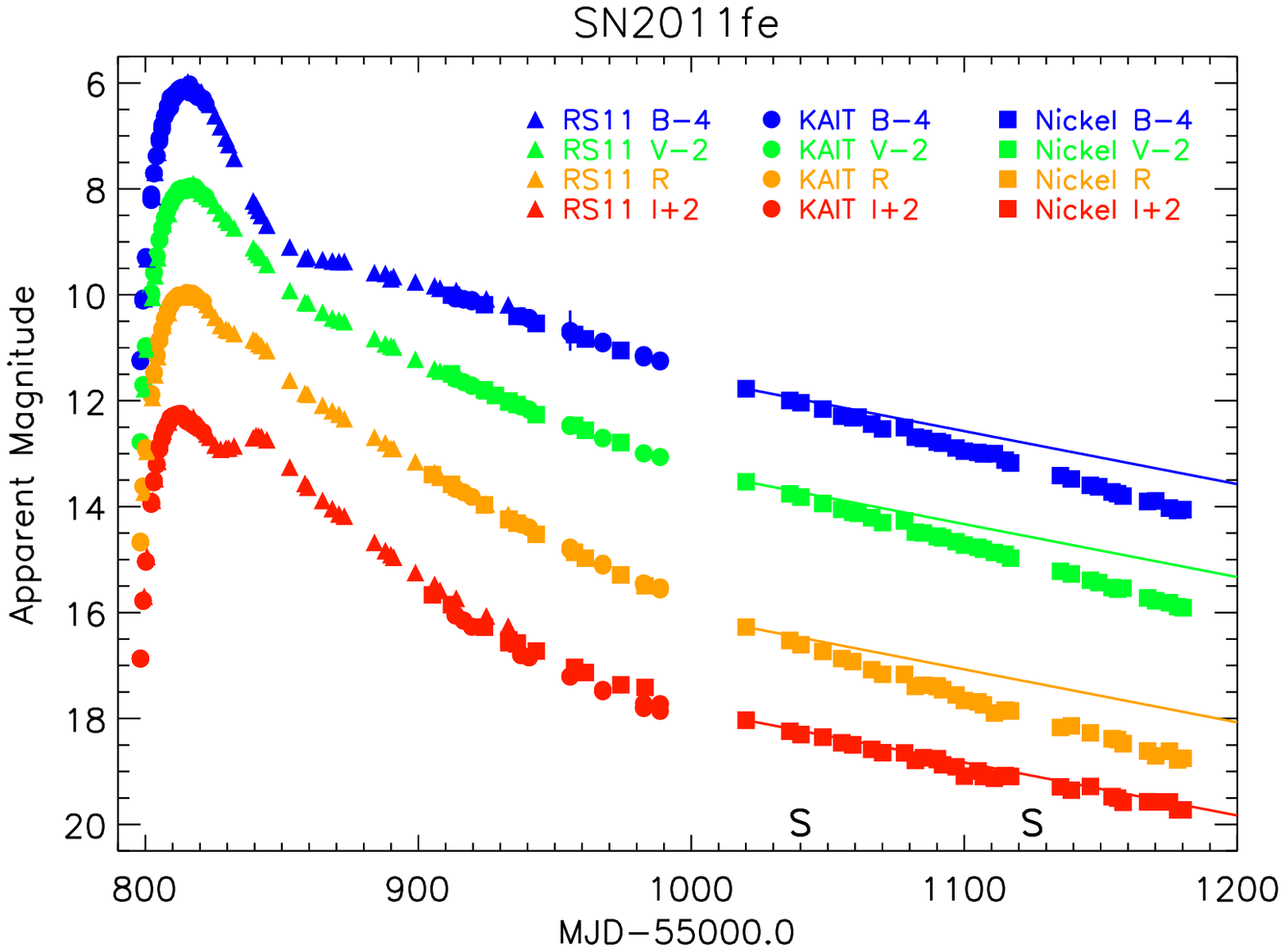}
\includegraphics[trim=1.2cm 0cm 0.2cm 0cm,clip=true,width=8.5cm]{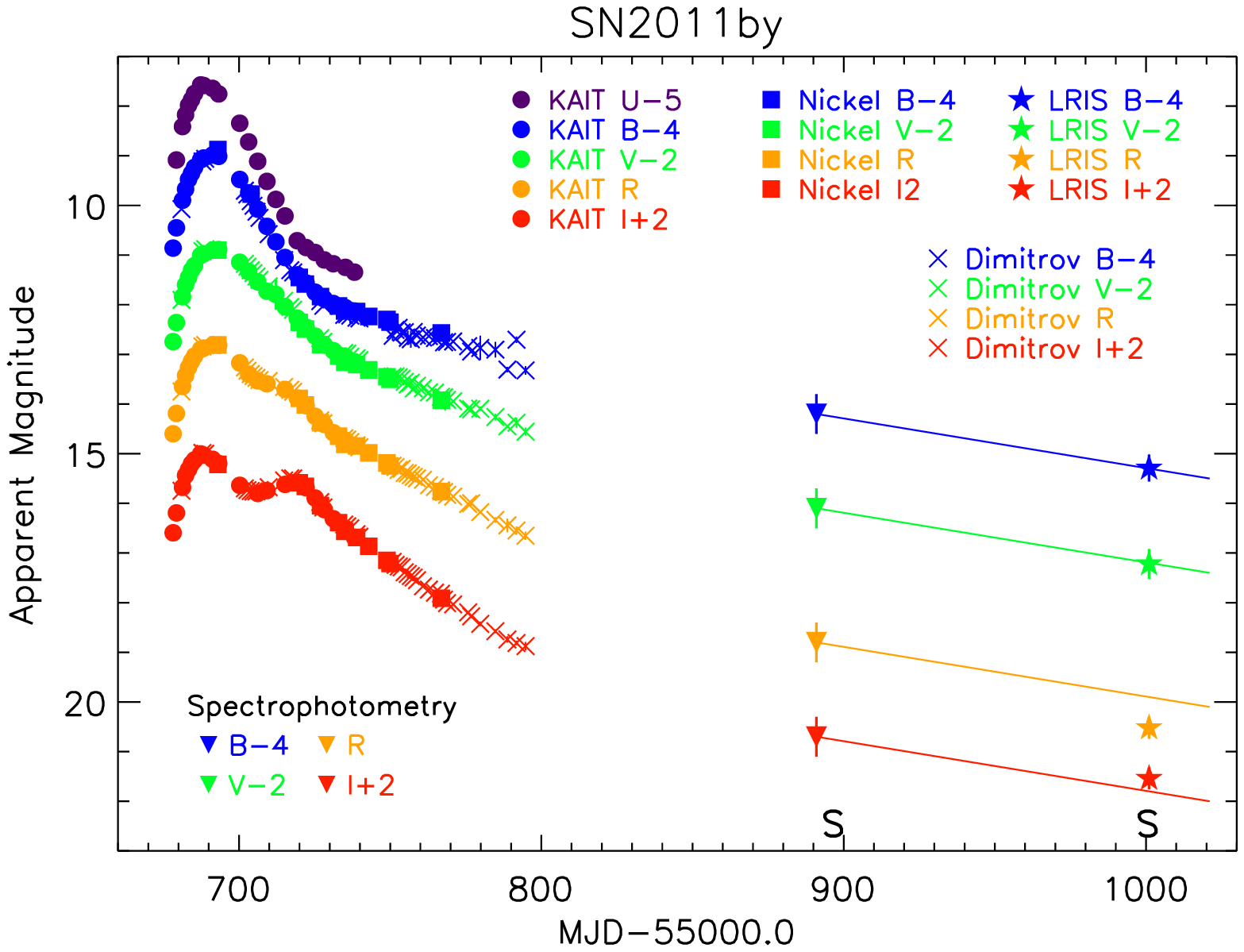}
\caption{Apparent magnitude $BVRI$ light curves of SN\,2011fe (left) and SN\,2011by (right). Photometry for SN\,2011fe includes data from KAIT (circles), the Nickel telescope (squares), and Richmond \& Smith 2011 (RS11, triangles). Photometry for SN\,2011by includes data from KAIT (circles), spectrophotometry from the +200 day LRIS nebular spectrum flux calibrated to luminance-filter photometry at +200 days (inverted triangles), LRIS imaging (five-point stars), and Dimitrov \& Kjurkchieva (2014, crosses). The ``S'' labels along the abscissa mark the times of nebular phase spectra. For both plots, a line representing cobalt decay of 0.01\,mag day$^{-1}$ has been extended from a light-curve point around day +200. \label{fig:lc}}
\end{center}
\end{figure*}

\subsection{Photometry}

We obtained photometry for SNe\,2011fe and 2011by with the 0.76\,m Katzman Automatic Imaging Telescope (KAIT; Filippenko et al. 2001) and the 1.0\,m Nickel telescope at Lick Observatory. We followed SN\,2011fe for a year after explosion, except for three months of the post-maximum decline when it was inaccessible. To complete the light curve we incorporate the published photometry of Richmond \& Smith (2012). We followed SN\,2011by in the $U$-band to $\sim40$ days, and in the $BVRI$ bands to $\sim75$ days past maximum brightness. We also obtained a single late-time photometric epoch for SN\,2011by in $BVRI$ with the Low Resolution Imaging Spectrometer (LRIS; Oke et al. 1995) on the 10\,m Keck-1 telescope. This photometry for SN\,2011by was previously published by Silverman et al. (2013). To extend the photometric coverage of SN\,2011by we include the published $BVRI$ photometry of Dimitrov \& Kjurkchieva (2014). 

All photometry has been reduced with standard methods and calibrated to the Vega magnitude system. Although we are incorporating a variety of sources, we do not need to derive or apply any filter transformations because we have KAIT and Nickel photometry for both SNe\,Ia at similar phases with which to make internally consistent comparisons of the colour evolution. Generally, all the photometry is in good agreement. Since we have only one set of late-time images for SN\,2011by, we have repeated our calibrations with several different star catalogs but find that all results agree within the measurement uncertainty (a few tenths of a magnitude).

To compensate for the lack of photometric coverage for SN\,2011by between days +100 and +300, we synthesise $BVRI$ magnitudes from the nebular phase spectrum at +200 days. First, we flux calibrate the +200 day spectrum to the luminance filter observation on 26 Nov. 2011 of $m_L = 18.3\pm0.3$\,mag posted by Joe Brimacombe on the Rochester SN page\footnote{http://www.rochesterastronomy.org/supernova.html}. The luminance filter transmission function, $T_L$, is similar to a combination of $BVR$ filters. We approximate it using the maximum value of the transmission function of $B$, $V$, and $R$ LRIS filters at each wavelength.  To estimate the systematic uncertainty introduced by our approximation we also define $T_L$ as a step function, where $T_L = 1.0$ for $4000 < \lambda < 7000$ \AA\ and $T_L=0$ for all other wavelengths. This represents an ``extreme" or ``maximum" transmission function estimate. We find that the difference in resulting magnitudes is $\sim0.1$ in all filters, and include this as a systematic uncertainty in the +200 spectrophotometry. We also include a relatively small additional uncertainty of $\sim0.02$\,mag from the $\sim5\%$ uncertainty in the relative flux scaling of the blue and red sides of LRIS (below and above $\lambda \approx 5500$ \AA) discussed in \S \ref{ssec:ananeb}. Our synthesised spectrophotometric data for MJD = 55891 (+200 days past maximum) is $B=18.2$, $V=18.1$, $R=18.8$, and $I=18.7$ ($\pm 0.4$) mag.

\subsection{Spectroscopy}

Photospheric NUV and optical spectra of SNe\,2011fe and 2011by were acquired with the Space Telescope Imaging Spectrograph (STIS) on $HST$ under programme GO-12298 (PI: R. Ellis), all of which are publicly available. Coverage in the NUV was obtained with the 52\arcsec\ x 0.2\arcsec\ slit and either the G230L grating, which covers $\lambda=1570$--3180 \AA\ with a dispersion of 1.58 \AA/pixel, or the G230LB grating, which covers $\lambda=1685$--3060 \AA\ with a dispersion of 1.35 \AA/pixel. Coverage in the optical was obtained with the 52\arcsec\ x 0.2\arcsec\ slit and the G430L grating, which covers $\lambda=2900$--5700 \AA\ with a dispersion of 2.73 \AA/pixel, and the G750L grating, which covers $\lambda=5240$--10,270 \AA\ with a dispersion of 4.92 \AA/pixel. All $HST$+STIS spectra used in this work are listed in Table \ref{tab:spec}. The $HST$+STIS spectra taken near maximum light for both objects were previously published by FK13, and those for the photospheric phase spectra of SN\,2011fe were previously published by Mazzali et al. (2014); we use their carefully reduced versions available at WISEREP\footnote{www.weizmann.ac.il/astrophysics/wiserep} (Yaron \& Gal-Yam 2012).

We also acquired spectra of SNe\,2011fe and 2011by at the Lick and Keck Observatories under the long-running Berkeley Supernova Ia Program (BSNIP; Silverman et al. 2012), as listed in Table \ref{tab:spec}. At Lick Observatory we used the Kast spectrograph (Miller \& Stone 1993) on the 3\,m Shane telescope. Coverage in the blue was obtained with the 600/4310 grism ($\lambda=3300$--5520 \AA, 1.02 \AA/pixel), and coverage in the red with the 300/7500 grating ($\lambda=5400$--10,400\AA, 4.60 \AA/pixel). At Keck Observatory we used LRIS to obtain the two nebular phase spectra of SN\,2011by. Coverage in the blue was obtained with the 600/4000 grism ($\lambda=3010$--5600 \AA, 0.63 \AA/pixel), and in the red with the 400/8500 grating ($\lambda=4500$--9000 \AA, 1.16 \AA/pixel). The BSNIP spectra were previously published by Silverman et al. (2013). For ground-based optical spectroscopy, the slit was generally aligned along the parallactic angle (Filippenko 1982) to minimize the effects of atmospheric dispersion (in addition, LRIS has an atmospheric dispersion corrector).

\section{Analysis} \label{sec:ana}

Here we present and analyse the photometric and spectroscopic observations of SNe\,2011fe and 2011by. This section is organised by data type, starting with the light curves in \S \ref{ssec:anaphot}, including the apparent late-time photometric excess of SN\,2011by in \S \ref{sssec:late}. The photospheric phase NUV-optical spectra are compared in \S \ref{ssec:anaspec}, where we demonstrate the extreme similarity in optical evolution, but the relatively depressed NUV flux of SN\,2011by. In \S \ref{ssec:ananeb} we present the optical nebular phase spectra, including a comparison with other nebular SNe\,Ia in \S \ref{sssec:nebcomp}, a careful examination of their residual spectra in \S \ref{sssec:nebresi}, and an evaluation of the flux, velocity, and width of the nebular emission features in \S \ref{sssec:nebeval}. The results of these analyses are interpreted with physical models in \S \ref{sec:disc}.

\subsection{Light Curves of SNe\,2011fe and 2011by}\label{ssec:anaphot}

\begin{figure}
\begin{center}
\includegraphics[width=8.5cm]{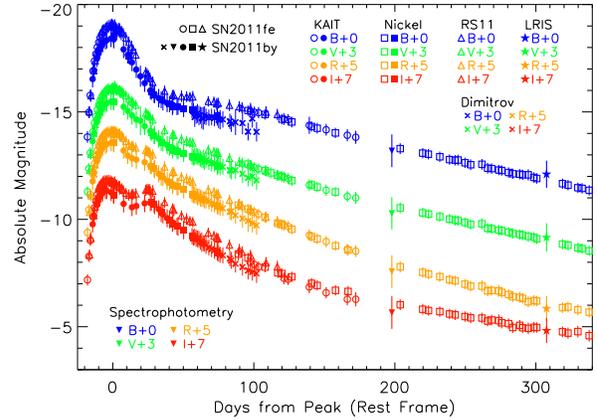}
\caption{Absolute magnitude $BVRI$ light curves of SN\,2011fe (open) and SN\,2011by (filled), as a function of rest-frame time from $B$-band peak magnitude, using the same symbol convention as in Figure \ref{fig:lc}. Photometry from Dimitrov \& Kjurkchieva (2014) is shown only after 50 days past peak for clarity. Error bars are mostly from the uncertainties in the host-galaxy distance modulus. The magnitude differences, $m_{\rm fe}-m_{\rm by}$, at peak light in $BVRI$ are (respectively) $-0.6$, $-0.7$, $-0.5$, and $-0.5$; at +300 days they appear to have decreased to (respectively) +0.2, +0.1, $-0.3$, and $-0.1$. \label{fig:lc_comp}}
\end{center}
\end{figure}

\begin{figure}
\begin{center}
\includegraphics[trim=0cm 0cm 0.5cm 0cm,clip=true,width=8.5cm]{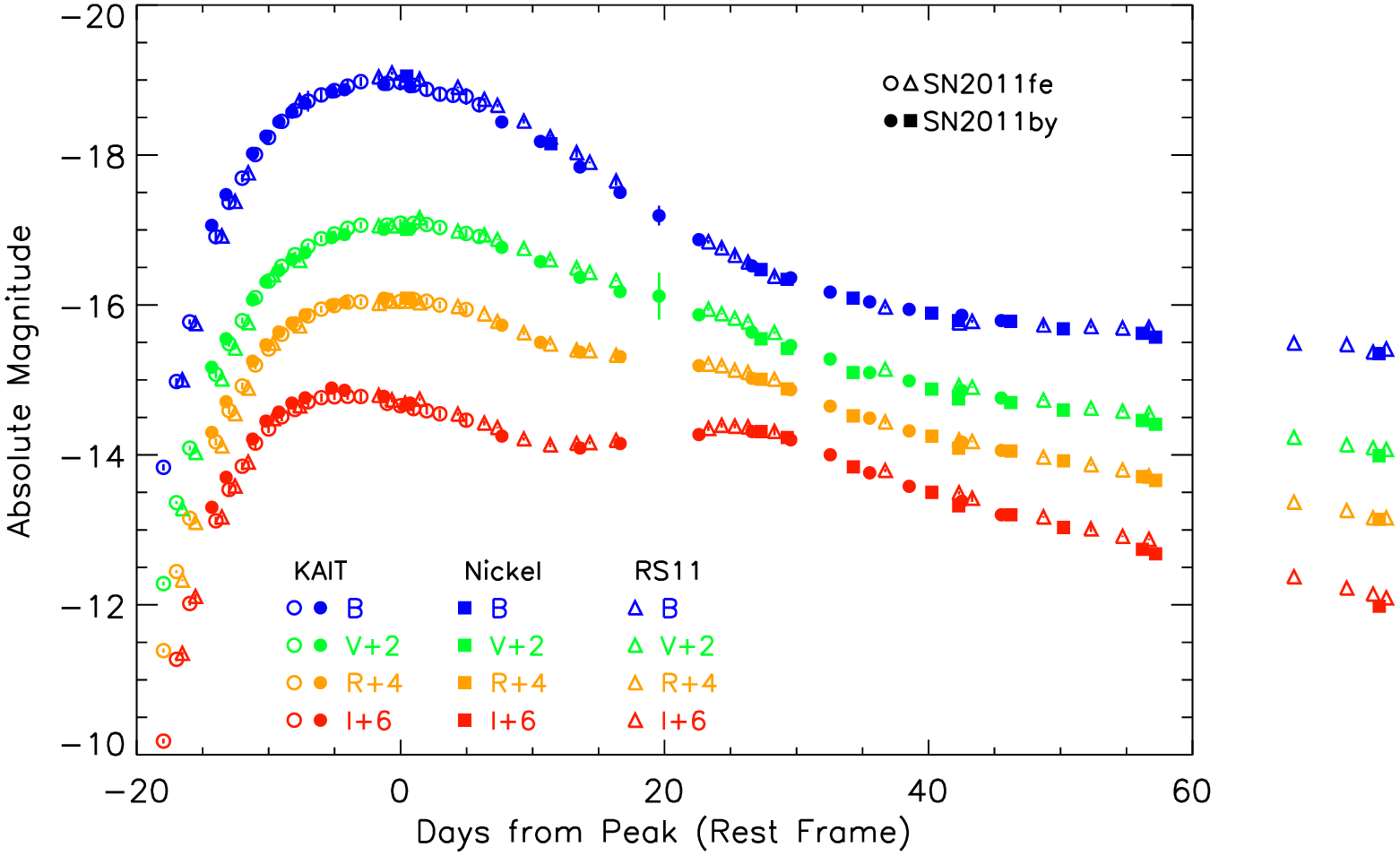}
\caption{To emphasise the similarity in light-curve shapes, we apply a constant to all absolute magnitudes of SN\,2011by (filled) to match its peak $B$-band magnitude to that of SN\,2011fe (open). \label{fig:lc_comp_max}}
\end{center}
\end{figure}

\begin{figure}
\begin{center}
\includegraphics[width=8.5cm]{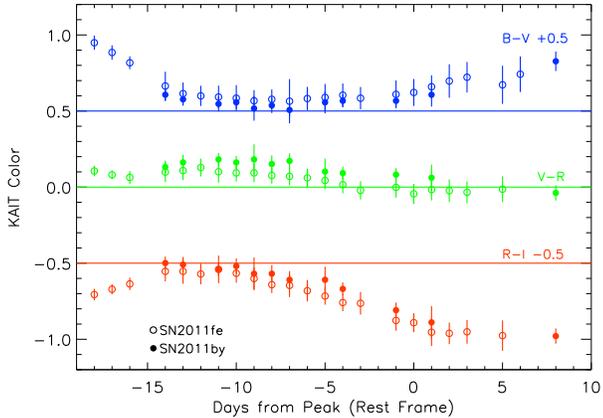}
\caption{Colour evolution over maximum brightness of SN\,2011fe (open) and SN\,2011by (filled) in KAIT $B-V$, $V-R$, and $R-I$ (magnitude units). Colours are offset on the ordinate for clarity, with horizontal lines marking zero for each colour. Error bars represent the measurement uncertainties. \label{fig:clr}}
\end{center}
\end{figure}

In Figure \ref{fig:lc} we display the optical light curves of SNe\,2011fe and 2011by in apparent magnitudes. We find that they are consistent with the previously published light-curve-fit parameters listed in Table \ref{tab:params}. In Figure \ref{fig:lc_comp} we show the optical light curves in absolute magnitudes, after applying the Galactic extinction values and distance moduli listed in \S \ref{sec:obs} (using the Tully-Fisher distance modulus for SN\,2011by). As established by FK13, SN\,2011by appears to be $\sim0.6$\,mag less luminous than SN\,2011fe at peak brightness, despite having a similar light-curve shape and decline rate, $\Delta m_{15}(B)$. According to the work of Phillips et al. (1999), the dispersion in peak brightness for unreddened SNe\,Ia is $\sigma=0.11$\,mag. After accounting for the slightly faster $\Delta m_{15}(B)$ of SN\,2011by, we find that it lies $\sim5\sigma$ from the established width-luminosity relation for SNe\,Ia (Phillips 1993).

FK13 used the work of Arnett (1982) to show that, using the Tully-Fisher distance modulus for SN\,2011by, SN\,2011fe formed $1.7_{-0.5}^{+0.7}$ times as much radioactive $^{56}$Ni as SN\,2011by. Pereira et al. (2013) determined that SN\,2011fe synthesised $0.53\pm0.11\,\rm M_{\odot}$ of $^{56}$Ni. Together, we surmise that SN\,2011by synthesised $0.31_{0.16}^{0.19}\,\rm M_{\odot}$ of $^{56}$Ni, which is $\sim0.2\,\rm M_{\odot}$ less than SN\,2011fe. These calculations rely on their having similar rise times (which they do; see Table \ref{tab:params}), and similar filter-to-bolometric luminosity corrections. Despite the depressed NUV flux for SN\,2011by, the latter assumption is reasonable because the bolometric contribution from the NUV flux is about an order of magnitude less than that from the optical flux.

The abnormally low peak luminosity, and the inferred low $^{56}$Ni mass, of SN\,2011by with respect to SN\,2011fe are difficult to reconcile with their $BVRI$ light-curve shapes and colours being well matched. To emphasise their similarity in shape, in Figure \ref{fig:lc_comp_max} we simply shift the absolute magnitudes of SN\,2011by by a constant in all filters, such that its $B$-band peak magnitude matches that of SN\,2011fe. This illustrates how either SN\,2011by lies $\sim5\sigma$ off the Phillips relation, or the Tully-Fisher distance to NGC 3972 is underestimated. In the latter case, SNe\,2011fe and 2011by may have reached similar peak absolute magnitudes, and synthesised similar masses of $^{56}$Ni. In \S \ref{ssec:anaphot} and \S \ref{ssec:ananeb} we demonstrate that their twin-like optical spectral evolution suggests that the two SNe\,Ia experienced physically similar explosions, which supports the latter hypothesis.

In Figure \ref{fig:lc_comp_max}, when matched to SN\,2011fe in the $B$ band, SN\,2011by appears slightly fainter in $V$ and brighter in $I$. To further examine this subtle distinction, in Figure \ref{fig:clr} we plot the colour evolution of SNe\,2011fe and 2011by in $B-V$, $V-R$, and $R-I$. For internal consistency, we use only the KAIT photometry which was all taken with the same filters and processed with the same pipeline and sequence of local standard stars. The colour discrepancies are $<0.1$\,mag, a flux difference of just 5\% in each band (i.e., not large enough to contradict their nearly identical optical spectra). The similar colours of SNe\,2011fe and 2011by indicate that the potential $\sim0.6$\,mag disparity cannot be explained by dust extinction. This conclusion is further supported by the fact that our own light-curve fits using MLCS2k2 (Jha, Riess \& Kirshner 2007) arrive at a similar distance modulus, regardless of whether the extinction-curve slope $R_V$ is held fixed at a Milky-Way value of 3.1 or allowed to vary. 

\vspace{-10pt}
\subsubsection{The Late-Time Decline of SN\,2011by}\label{sssec:late}

Given that the rise and decline rates of SN\,2011fe and 2011by are very similar up to 100 days past maximum light, it is reasonable to expect that they continue to experience a similar decline rate, and that at 200 days past maximum they would exhibit the same difference in magnitude that we see at early times. For example, in the case where we assume the Tully-Fisher distance modulus for SN\,2011by, we expect it to be $\sim0.6$\,mag fainter than SN\,2011fe at late times (i.e., the offset applied to match light-curve peaks in Figure \ref{fig:lc_comp_max}). However, as seen in Figure \ref{fig:lc_comp}, by +200 days it is apparent that SN\,2011by has experienced a slower decline, as it is within $\sim0.1$\,mag of SN\,2011fe in the $B$ band at +200 and +300 days. In the scenario where the Tully-Fisher distance is an underestimate and both SNe\,Ia reached similar peak absolute magnitudes, SN\,2011by is more luminous than SN\,2011fe at late times. Regardless of distance, SN\,2011by declined more slowly than expected at late times. What could be the physical cause of this difference in late-time decline rate between two otherwise similar SNe\,Ia?

At these late epochs the light curves are powered by the decay of $^{56}$Co to $^{56}$Fe, for which the luminosity decline rate is 0.01\,mag day$^{-1}$. Owing to incomplete trapping of gamma rays and positrons, the bolometric decline is faster than this, but the $I$ band declines more slowly because of specific lines in that band. In Figure \ref{fig:lc} we represent the $^{56}$Co decline as solid lines extending from +200 days past maximum for SNe\,2011fe and 2011by. We find that the late-time decline of SN\,2011fe is typical of SNe\,Ia, but the late-time photometry for SN\,2011by exhibits a slower rate of decline in $B$, $V$, and $I$. When does this slower decline begin for SN\,2011by? The photometry from Dimitrov \& Kjurkchieva (2014), shown after 50 days from peak in Figure \ref{fig:lc_comp}, is noisy in the $B$ band but appears to have a slightly slower decline rate. In $V$, $R$, and especially $I$, the gap between SNe\,2011fe and 2011by is steadily closing. It appears that the change in the decline rate for SN\,2011by starts at a phase $<100$ days, but owing to its poor coverage and large uncertainty at +200 days we cannot constrain exactly when or how its light-curve decline rate deviates from that of SN\,2011fe. 

Here we consider three potential physical causes of a late-time luminosity excess that are independent of the SN\,Ia explosion: host-galaxy contamination, a light echo, and interaction with circumstellar material (CSM). 

(1) Could the slower decline of SN\,2011by be caused by relatively larger host-galaxy contamination at late times? The line of sight to SN\,2011by is closer to the core of its host galaxy than that of SN\,2011fe, and may suffer some contamination from host-galaxy light. We have inspected the images and estimate that the host-galaxy light is not more than 15\% of the SN flux at late times, which is only 0.15\,mag (within our error bars). Furthermore, host contribution would be higher in the red filters than in the blue, but the difference we see is the opposite. 

(2) Could the excess luminosity of SN\,2011by at +300 days be caused by a light echo? The effect of reflected light would be stronger in the blue than in the red filters, as observed. The creation of a light echo implies dust within a light year, and the best candidates for light echos are SNe that exhibit dust extinction and reddening at earlier times, but as previously established there is little to no reddening for SN\,2011by. It could be that the reflecting material is oriented in a face-on disc, or is a dust sheet behind the SN\,Ia, and does not interfere with our line of sight to SN\,2011by. However, if an echo were making the observed magnitude brighter by $\sim0.5$\,mag, this would imply that approximately half of the flux in the nebular phase spectrum should be from an echo of the maximum-light emission. As presented below, the nebular phase spectra of SNe\,2011fe and 2011by are remarkably similar. Since we do not see any evidence for earlier-phase features in the nebular spectrum of SN\,2011by, we conclude that this late-time excess is not dominated by a light echo.

(3) Could interaction with CSM around SN\,2011by enhance the luminosity at late times? This hypothesis faces the same arguments as the light-echo hypothesis. The material would have to be even closer to the progenitor system, and is therefore more likely to cause an effect on the early-time spectra (e.g., narrow absorptions lines, unless the orientation is a face-on disc). We would also expect to see a contribution from this interaction in the nebular phase spectra (e.g., narrow emission lines).

The slower decline and excess luminosity of SN\,2011by at late times does not appear to be caused by host-galaxy contamination, a light echo, or CSM interaction, and furthermore it is not dependent on assumptions about the distance modulus. The remaining explanations are physical differences in the explosion and/or nucleosynthetic products of SN\,2011by, which we discuss in \S \ref{ssec:discLC}.

\begin{figure}
\begin{center}
\includegraphics[trim=0.3cm 1cm 1cm 2cm,clip=true,width=8.5cm]{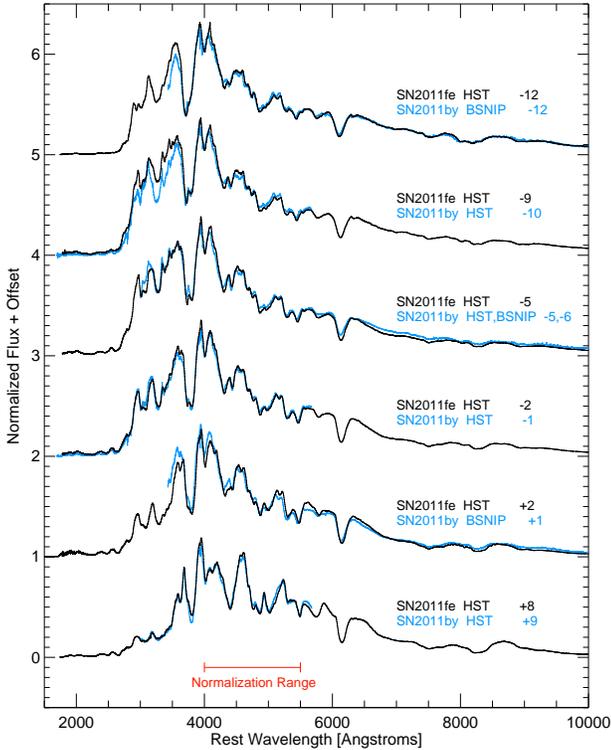}
\caption{Paired NUV-optical photospheric phase spectra of SNe\,2011fe (black) and 2011by (blue) show similar evolution of their features. Spectra are from the $HST$ and BSNIP surveys, as listed in Table \ref{tab:spec}. Only spectra that are matched to within 1 day are included. \label{fig:optevol}}
\end{center}
\end{figure}

\begin{figure}
\begin{center}
\includegraphics[trim=0cm 1cm 1cm 2cm,clip=true,width=8cm]{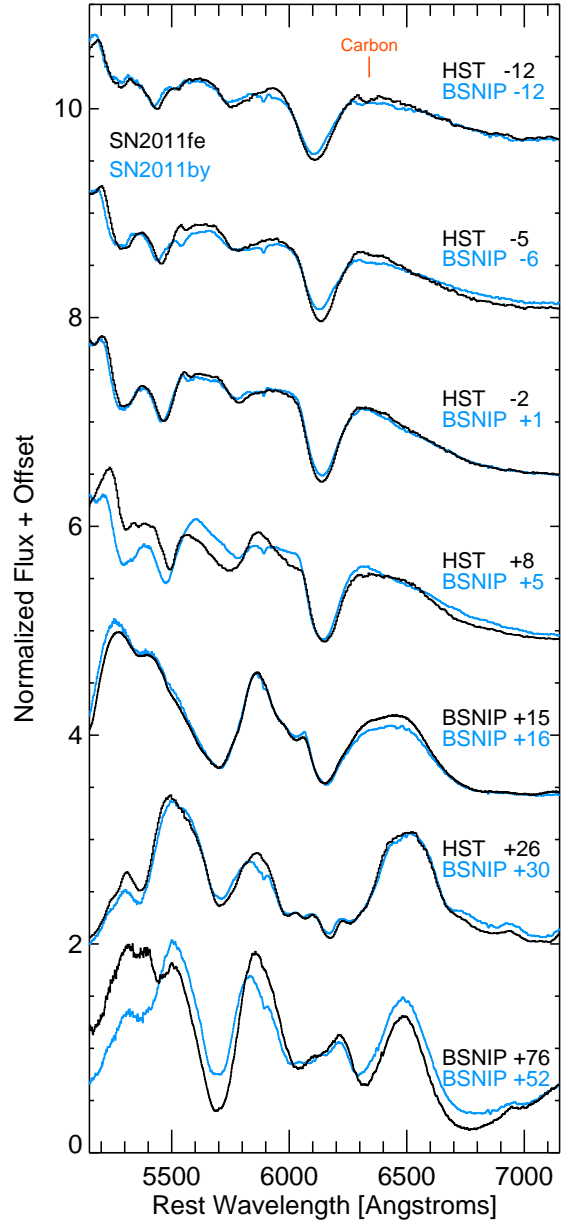}
\caption{Temporal evolution of the \ion{Si}{II} $\lambda$6355 line from BSNIP and $HST$ photospheric spectra for SNe\,2011fe (black) and 2011by (blue). Spectra have been normalised, matched by phase, and offset along the ordinate for clarity. An orange tick mark notes the position of carbon. \label{fig:Si6150}}
\end{center}
\end{figure}

\begin{figure}
\includegraphics[width=8.5cm]{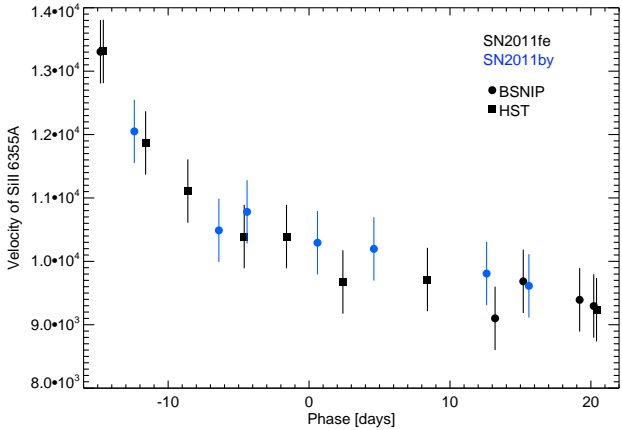}
\caption{The velocity of \ion{Si}{II} $\lambda$6355 in photospheric spectra of SN\,2011fe (black) and SN\,2011by (blue) from BSNIP (circles) and $HST$ (squares) until $\sim20$ days past maximum brightness. The \ion{Si}{II} $\lambda$6355 velocity evolution of SNe\,2011fe and 2011by are quite similar. Velocities are not Gaussian fits, but simply measured from the minima, with a $\pm10$ \AA\ uncertainty ($\pm500\,\rm km\ s^{-1}$). \label{fig:vel_Si6150}}
\end{figure}

\begin{figure*}
\begin{center}
\includegraphics[trim=0cm 0cm 0cm 0cm,clip=true,width=8cm]{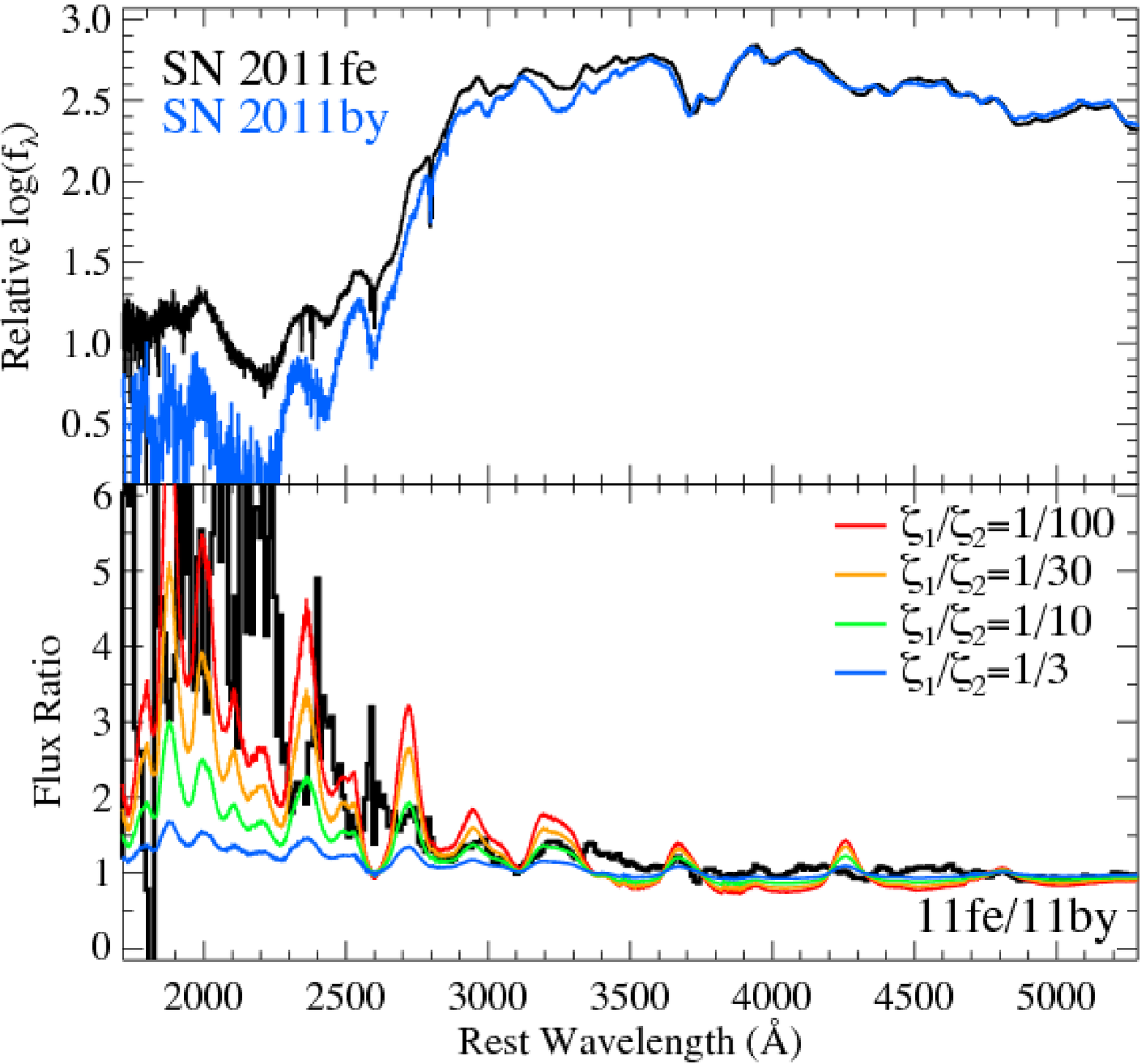}
\includegraphics[trim=0cm 0cm 0cm 0cm,clip=true,width=8cm]{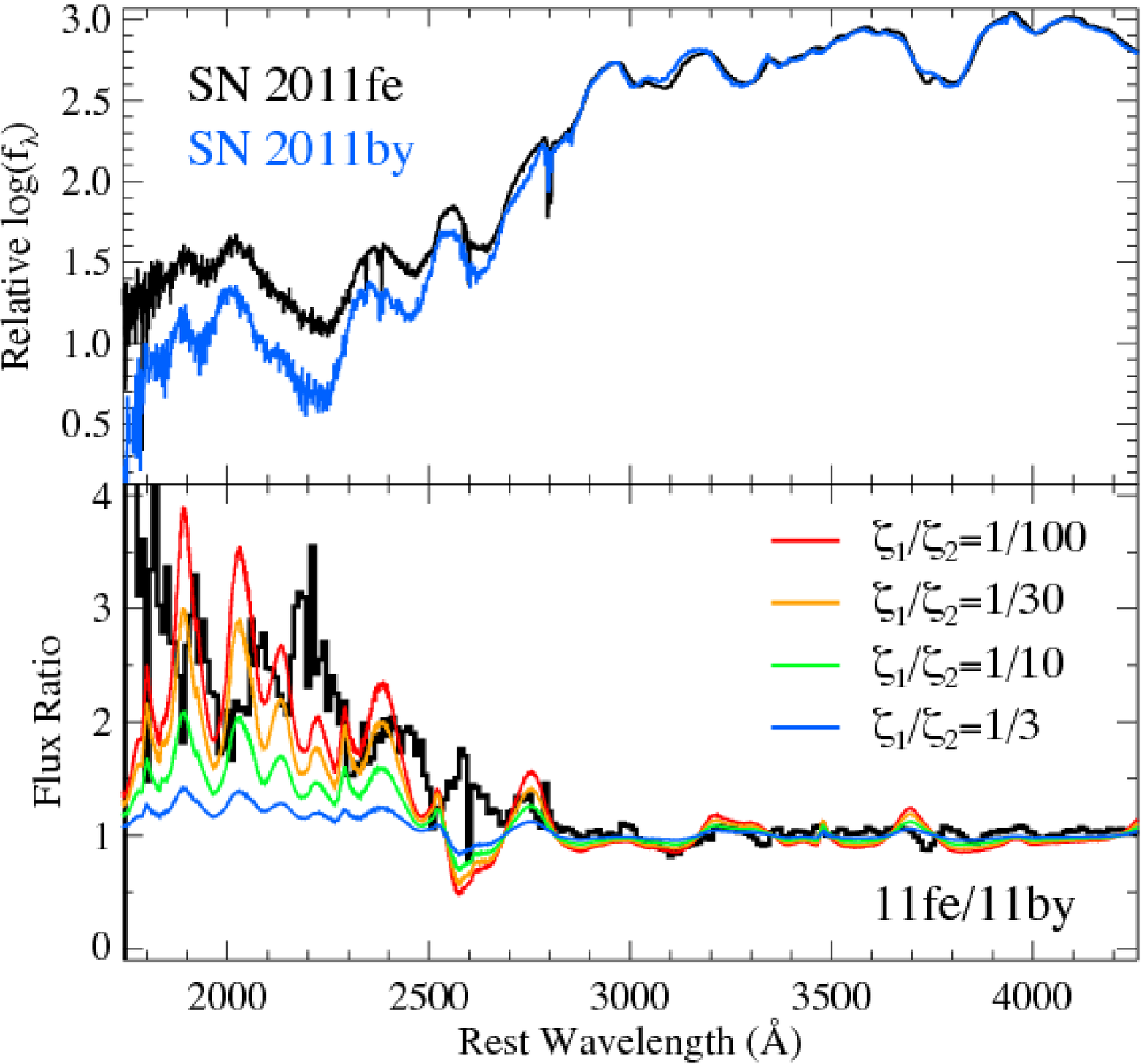}
\caption{Comparison of photospheric phase NUV-optical spectra of SN\,2011fe (black) and SN\,2011by (blue), for pre-maximum phases $-9$ and $-10$ days (left), and near-maximum phases $-2$ and $-1$ days (right), for SNe\,2011fe and 2011by (respectively). The near-maximum phases were shown previously by FK13; we have included them for comparison. The top panels display the spectra, flux scaled to match at optical wavelengths, and plotted on a logarithmic axis to emphasis the discrepancy at short wavelengths where the flux is low. Bottom panels show the ratio of fluxes at each wavelength in black; the blue, green, orange, and red lines represent the predicted flux ratio from an increasing difference in progenitor metallicity at that phase (Lentz et al. 2000). \label{fig:specearly}}
\end{center}
\end{figure*}

\subsection{Photospheric Phase Spectra}\label{ssec:anaspec}

Around maximum light, the SN\,Ia photosphere is still relatively close to the original surface and the resulting spectrum is a blackbody with absorption lines from high-velocity material in the outer layers (e.g., silicon, sulfur, magnesium, calcium, and oxygen). In Figure \ref{fig:optevol}, we compare the NUV-optical spectral evolution of SNe\,2011fe and 2011by during the photospheric phase. These spectra come from the $HST$ and BSNIP surveys, as listed in Table \ref{tab:spec}. They have been deredshifted and dereddened using the host-galaxy recession velocities and line-of-sight Milky Way extinction values listed in \S \ref{sec:obs}, and have been flux normalised in the optical. Only spectra that are phase matched to within 1 day are included, and after +10 days the phases of the spectral observations are not well matched for such a comparison. The spectra exhibit very similar evolution, and both SNe\,Ia are spectroscopically normal\footnote{Note there is a flux-calibration problem with the blue side of the SN\,2011by BSNIP spectrum at +1 days. This spectrum is not used in any quantitative analysis.}.

As the explosion evolves, the photosphere recedes into the expanding ejecta and the \ion{Si}{II} $\lambda$6355 line is created by lower, slower-moving layers; in this way, the velocity gradient of \ion{Si}{II} $\lambda$6355 is directly related to the physical parameters of the explosion. In Figure \ref{fig:Si6150} we show the evolution of \ion{Si}{II} $\lambda$6355 from early to late times, with as close phase matching as possible (the post-maximum spectra were taken less frequently and the matching is not as close). For a more continuous evaluation, in Figure \ref{fig:vel_Si6150} we plot the \ion{Si}{II} $\lambda$6355 line velocity over time until 20 days past peak light. SNe\,2011fe and 2011by are classified as members of the ``low velocity gradient" (LVG) class (Benetti et al. 2005; Parrent et al. 2012; Silverman et al. 2013), and the evolution of \ion{Si}{II} $\lambda$6355 is remarkably similar between them. Additionally, they both show evidence of carbon as notches on the red side of the \ion{Si}{II} $\lambda$6355 line in the pre-maximum light spectra (see Figure \ref{fig:Si6150}), and we have confirmed the presence of \ion{C}{II} with SYN++ (Thomas et al. 2011) fits for the first two optical spectra of SN\,2011by. This is commonly interpreted as unburned progenitor material, and the fact that it persists to $\sim-5$ days in both SNe\,2011fe and 2011by suggests that it is present to a comparable depth in each SN. These parallels in the optical spectral evolution of SNe\,2011fe and 2011by indicate that they experienced very similar physical explosions. 

In the scenario where SN\,2011fe synthesised $1.7_{-0.5}^{+0.7}$ times as much radioactive $^{56}$Ni as SN\,2011by, if we assume the standard interpretation that SN\,Ia explosions have a fixed mass budget $\lesssim1.37\, \rm M_{\odot}$, then SN\,2011fe should have formed smaller amounts of stable nucleosynthetic products: $^{56}$Fe, $^{58}$Ni, and intermediate-mass elements (IMEs; e.g., silicon). The amount of IMEs can be estimated from the highest and lowest velocity measurements of \ion{Si}{II} $\lambda$6355 in the expanding photosphere by using velocity is a proxy for radius. If these two SNe\,Ia formed significantly different amounts of IMEs, this fact would manifest as different silicon line velocity gradients. Instead, Figures \ref{fig:Si6150} and \ref{fig:vel_Si6150} show that SNe\,2011fe and 2011by are well matched in this respect.

In contrast to their outstanding optical similarity, the spectra of SNe\,2011fe and 2011by are different at NUV wavelengths. This distinction was first presented and discussed for spectra near maximum light by FK13 (the $-2$ and $-1$ day spectra of SNe\,2011fe and 2011by). Here we extend this comparison to earlier times (the $-9$ and $-10$ day spectra of SNe\,2011fe and 2011by), and show both phases for comparison in Figure \ref{fig:specearly}. The top panels display the spectra, which have been flux scaled to match at optical wavelengths. The bottom panels show the ratio of the fluxes ($f_{\rm fe}/f_{\rm by}$) along with the flux ratios predicted for varying progenitor metallicity from Lentz et al. (2000). Compared to maximum-light spectra, the pre-maximum spectra exhibit a larger NUV flux discrepancy that extends to redder wavelengths. This is consistent with the model predictions of Lentz et al. (2000). The physical implications of this NUV flux discrepancy for these two otherwise twin supernovae is discussed in detail in \S \ref{sec:disc}.

\subsection{Nebular Spectra}\label{ssec:ananeb}

After $\sim100$ days post-maximum light a SN\,Ia enters its nebular phase, when the material is optically thin and photons from all internal regions escape, and a full accounting of the nucleosynthetic materials --- including those from the deepest regions of the SN --- can be made. The nebular phase spectrum is dominated by forbidden emission lines of nickel, cobalt, and iron, but before $\sim200$ days post-maximum these lines are typically too blended to be used as diagnostics of the explosion. The half-life of radioactive $^{56}$Ni is 6 days, and by +200 days any nebular nickel lines are created by the stable $^{58}$Ni synthesised in the explosion. In the decay chain $^{56}$Ni $\rightarrow$ $^{56}$Co $\rightarrow$ $^{56}$Fe, the half-life of radioactive $^{56}$Co is 77 days, so $\sim16\%$ and $\sim7\%$ remain by +200 and +300 days. At these times, the iron lines are generated by the combination of stable $^{58}$Fe synthesised in the explosion, and the decay product $^{56}$Fe. Our two epochs of nebular phase spectra of SN\,2011by were obtained at +206 and +310 days past peak brightness. For comparison with SN\,2011fe, we use the closest phase-matched spectra, taken at +224 and +309 days past peak brightness. Henceforth, the two nebular phase epochs will be described simply as +200 and +300 days.

In the following sections we compare with the nebular phase spectra of other SNe\,Ia, and use residual spectra and Gaussian line fits to make a careful assessment of any subtle distinguishing features between the nebular spectra of SNe\,2011fe and 2011by. We demonstrate that the nebular optical phase spectra of SNe\,2011fe and 2011by are even more twin-like than their photospheric phase spectra, and argue that this suggests that they synthesised similar ratios of nucleosynthetic products.

\subsubsection{Comparison with Other Normal SNe\,Ia}\label{sssec:nebcomp}

\begin{figure*}
\includegraphics[width=15cm]{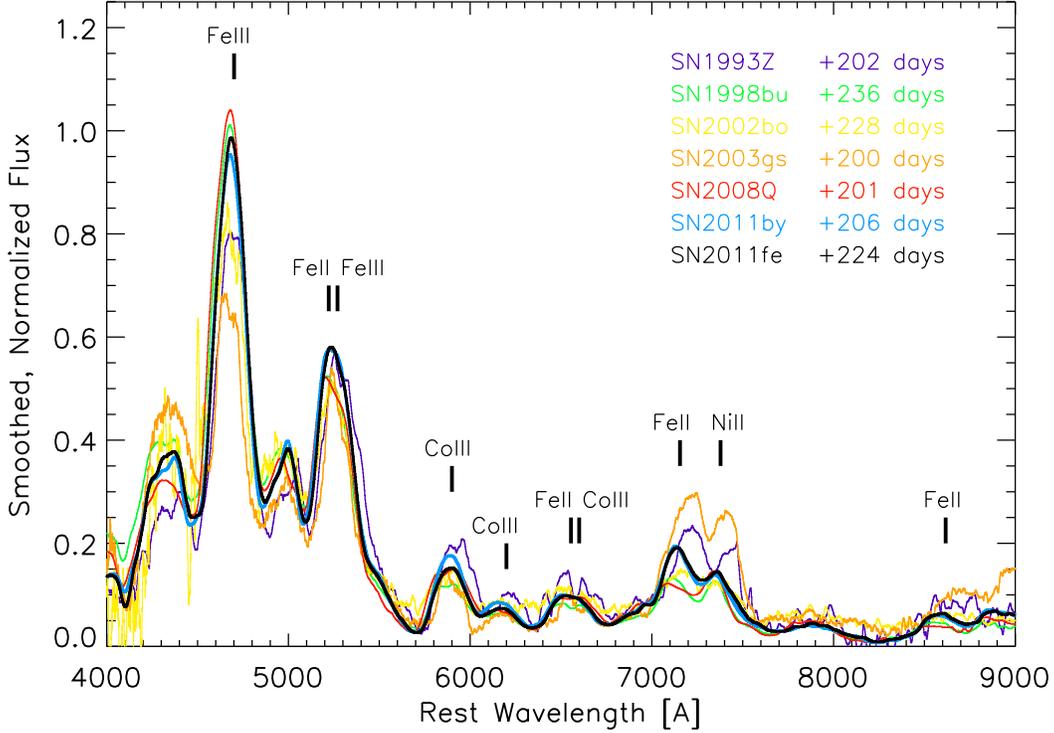}
\caption{The nebular spectra ($f_\lambda$) of SN\,2011fe (black) and SN\,2011by (blue) at +200 days past maximum light along with the nebular phase spectra of other SNe\,Ia at similar phases. For comparison purposes, all spectra have been smoothed, as well as normalised to have equal areas between 4000 and 9000 \AA. The strongest forbidden emission lines are labeled with their dominant species. Details of our comparison sample are described in the text. \label{fig:nebcomp}}
\end{figure*}

\begin{table*}
\begin{center}
\begin{minipage}[bp]{5.2in}
\setlength{\tabcolsep}{2.pt}
\caption{Comparison SNe\,Ia for Nebular Spectra in Figure \ref{fig:nebcomp} \label{tab:nebcomp}}
\centering
  \begin{tabular}[bp]{@{}cccccc@{}} 
  \hline
  \hline
SN\,Ia Name & Peak Brightness & $\Delta m_{15}(B)$ & $\dot{v}$(\ion{Si}{II})  & $E(B-V)$ & Sources\footnote{B=Benetti et al. 2004; C=Cappellaro et al. 2001; G10=Ganeshalingam et al. 2010; H=Hernandez et al. 2000; F=Foley \& Kirshner 2013; K=Krisciunas et al. 2009; R=Richmond \& Smith 2012; S12=Silverman et al. 2012; S12b=Silverman et al. 2012b; S13=Silverman et al. 2013} \\
\hline
SN\,1993Z  & --                   & --            & --  & 0.041 & S12, S12b, S13 \\
SN\,1998bu & $M_V=-19.37\pm0.23$  & $1.05\pm0.03$ & LVG & 0.368 & H, G10, S12, S12b, S13, C \\
SN\,2002bo & $M_B=-19.1\pm0.2$    & $1.15\pm0.04$ & HVG & 0.356 & B, G10, S12, S12b, S13 \\
SN\,2003gs & $M_B=-17.94\pm0.29$  & $1.83\pm0.02$ & --  & 0.065 & K, G10, S12, S12b, S13 \\
SN\,2008Q  & $M_B=-18.7\pm0.4$    & $1.25\pm0.08$ & LVG & 0.133 & G10, S12, S12b, S13 \\
SN\,2011by & $M_V=-18.47\pm0.36$  & $1.14\pm0.03$ & LVG & 0.012 & F, S12, S12b, S13 \\
SN\,2011fe & $M_V=-19.07\pm0.05$  & $1.21\pm0.03$ & LVG & 0.008 & F, R, S12, S12b, S13 \\
\hline
\end{tabular}
\vspace{-20pt}
\end{minipage} \end{center}
\end{table*}

Generally, normal SNe\,Ia are as homogeneous in their nebular spectra as they are in the photospheric phase, although there is diversity in the exhibited line velocity and strength (e.g., Fig. 2 of Mazzali et al. 2011; Maeda et al. 2010). In Figure \ref{fig:nebcomp} we show the +200 day spectra of SNe\,2011fe and 2011by along with five nebular spectra in the BSNIP database that were taken at similar phases with comparably good signal-to-noise ratio (Silverman et al. 2013). All spectra have been deredshifted and corrected for Galactic and (when applicable) host-galaxy extinction. These spectra are all of normal SNe\,Ia, with representation from those that were more and less luminous, and/or had higher \ion{Si}{II} velocity gradients, than SNe\,2011fe and 2011by. We list the relevant information for each in Table \ref{tab:nebcomp}. Note that in Figure \ref{fig:nebcomp} we use the original reduced spectra of SNe\,2011fe and 2011by, not the residual-minimised versions described by \S \ref{sssec:nebresi}. 

From this diverse cross section of normal SNe\,Ia at late times, Figure \ref{fig:nebcomp} shows that no other two spectra in our sample are as similar as SNe\,2011fe and 2011by are to each other --- they have continued to be twin-like at late times. This nebular twinness suggests that they were intrinsically similar explosions and produced similar amounts of nucleosynthetic materials. Here we briefly discuss each comparison spectrum individually.

\medskip
\noindent
$\bullet$ SN\,1993Z (purple) was discovered and classified as a SN\,Ia at $\sim4$ weeks after peak brightness (Treffers et al. 1993). As no near-maximum data exist, the light-curve decline rate and \ion{Si}{II} velocity gradient are not known, and no SN\,Ia subtype can be assigned.

\noindent
$\bullet$ SN\,1998bu (green) is a normal SN\,Ia with a low \ion{Si}{II} velocity gradient and substantial extinction from its host galaxy, but was intrinsically more luminous than SN\,2011fe and likely created more $^{56}$Ni (Hernandez et al. 2000). The +200 day nebular phase spectrum of SN\,1998bu is also typical of SNe\,Ia (Silverman et al. 2013), although at phases later than +500 days a light echo is clearly detected (Cappellaro et al. 2001). In SN\,1998bu at +230 days, the relatively higher flux in the blue iron lines and lower fluxes in the cobalt and nickel lines, compared to SNe\,2011fe and 2011by, is consistent with the higher $^{56}$Ni mass, but it could also be an early contribution from the light echo at +500 days.

\noindent
$\bullet$ SN\,2002bo (yellow) has a similar $\Delta m_{15}(B)$ decline rate as SNe\,2011fe and 2011by, and exhibited mostly normal SN\,Ia characteristics except for a redder $B-V$ colour at peak (that can be attributed to host extinction), and in contrast to our pair, exhibited a high \ion{Si}{II} velocity gradient (Benetti et al. 2004; Silverman et al. 2013).

\noindent
$\bullet$ SN\,2003gs (orange) declined rapidly and was optically subluminous, with a peculiar red spectrum similar to that of other subluminous SNe\,Ia, with no evidence of host extinction (Evans et al. 2003; Matheson \& Suntzeff 2003; Gonzalez, Morrell \& Hamuy 2003; Krisciunas et al. 2009; Silverman et al. 2013). SN\,2003gs likely produced significantly less $^{56}$Ni than SN\,2011fe, which explains the lower flux in the [\ion{Fe}{III}] $\lambda\approx 4700$ \AA\ line at late times. SN\,2003gs also appears to have synthesised relatively more stable nickel.

\noindent
$\bullet$ SN\,2008Q (red) has a decline rate of $\Delta m_{15}(B)=1.25$\,mag, a bit larger than those of SNe\,2011fe and 2011by, but it is otherwise similar around peak brightness and is also classified as having a low \ion{Si}{II} velocity gradient (Milne et al. 2013; Silverman et al. 2013). Like SN\,2011by and SN\,2011fe, SN\,2008Q is classified as NUV-blue by Milne et al. (2013). 

\vspace{-10pt}
\subsubsection{Residual Spectra}\label{sssec:nebresi}

\begin{figure*}
\includegraphics[trim=0cm 1.69cm 0cm 0cm,clip=true,width=8.5cm]{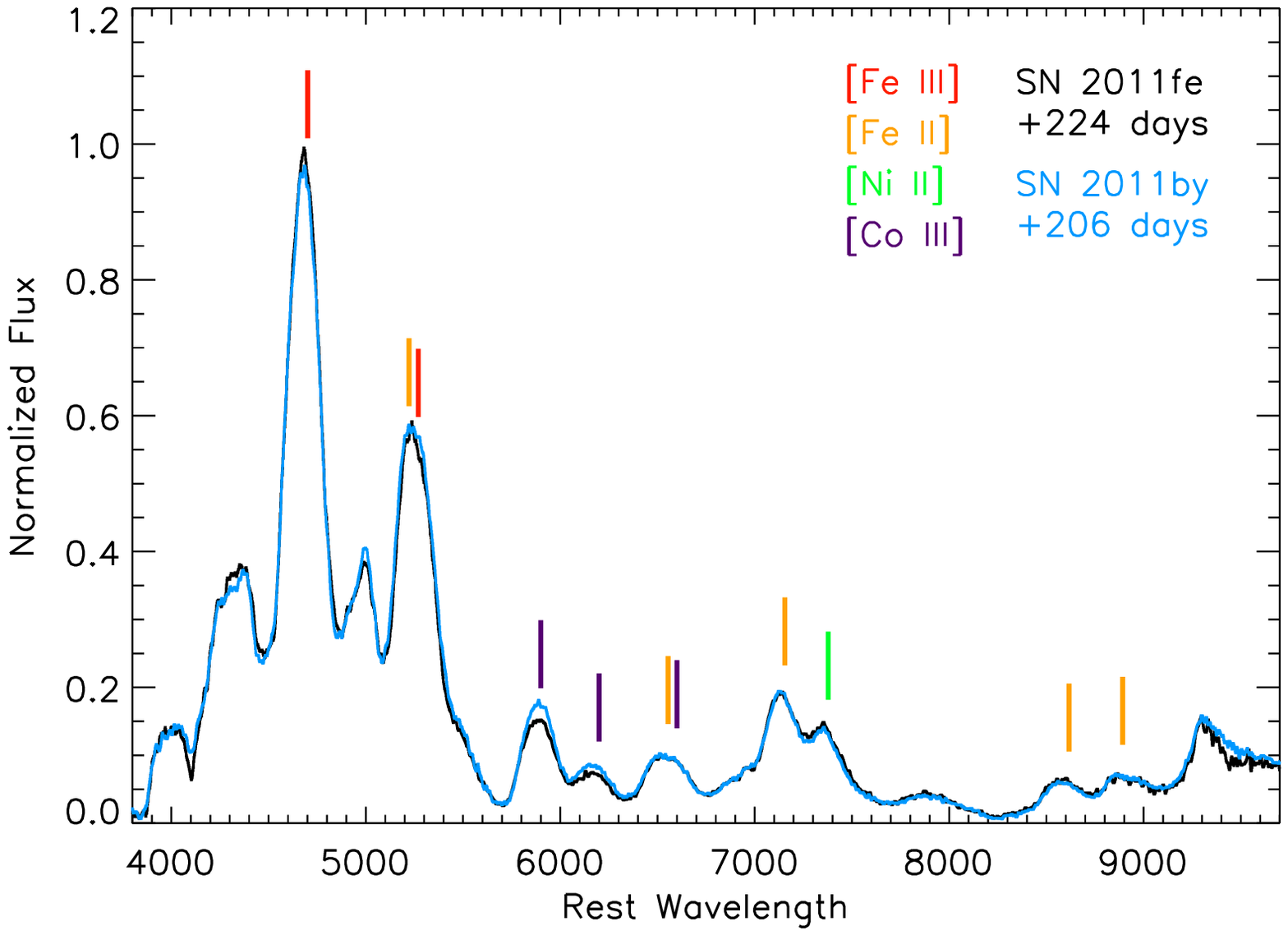}
\includegraphics[trim=0cm 1.69cm 0cm 0cm,clip=true,width=8.5cm]{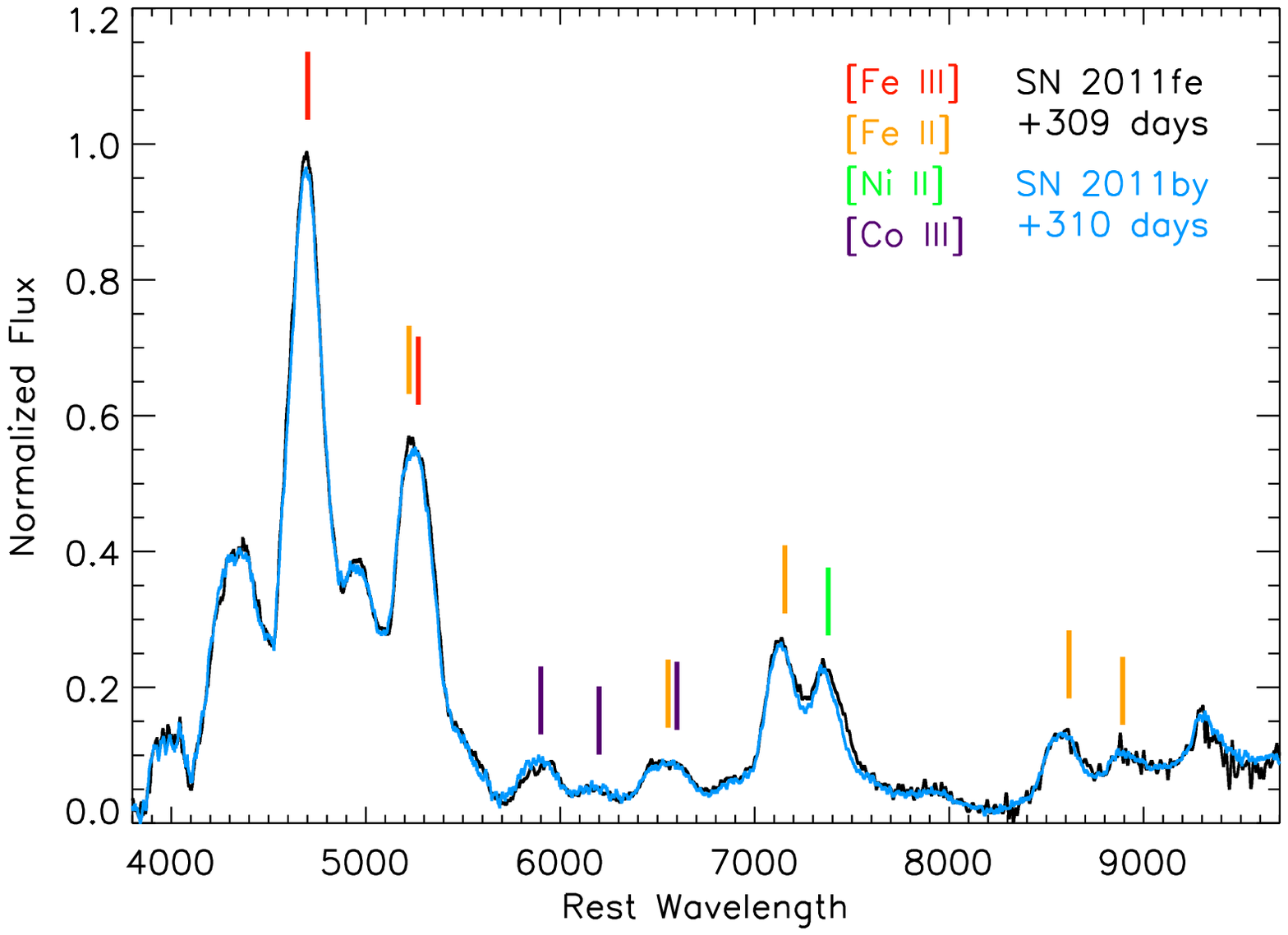}
\vskip -0.15em
\includegraphics[trim=0cm 0cm 0cm 0.84cm,clip=true,width=8.5cm]{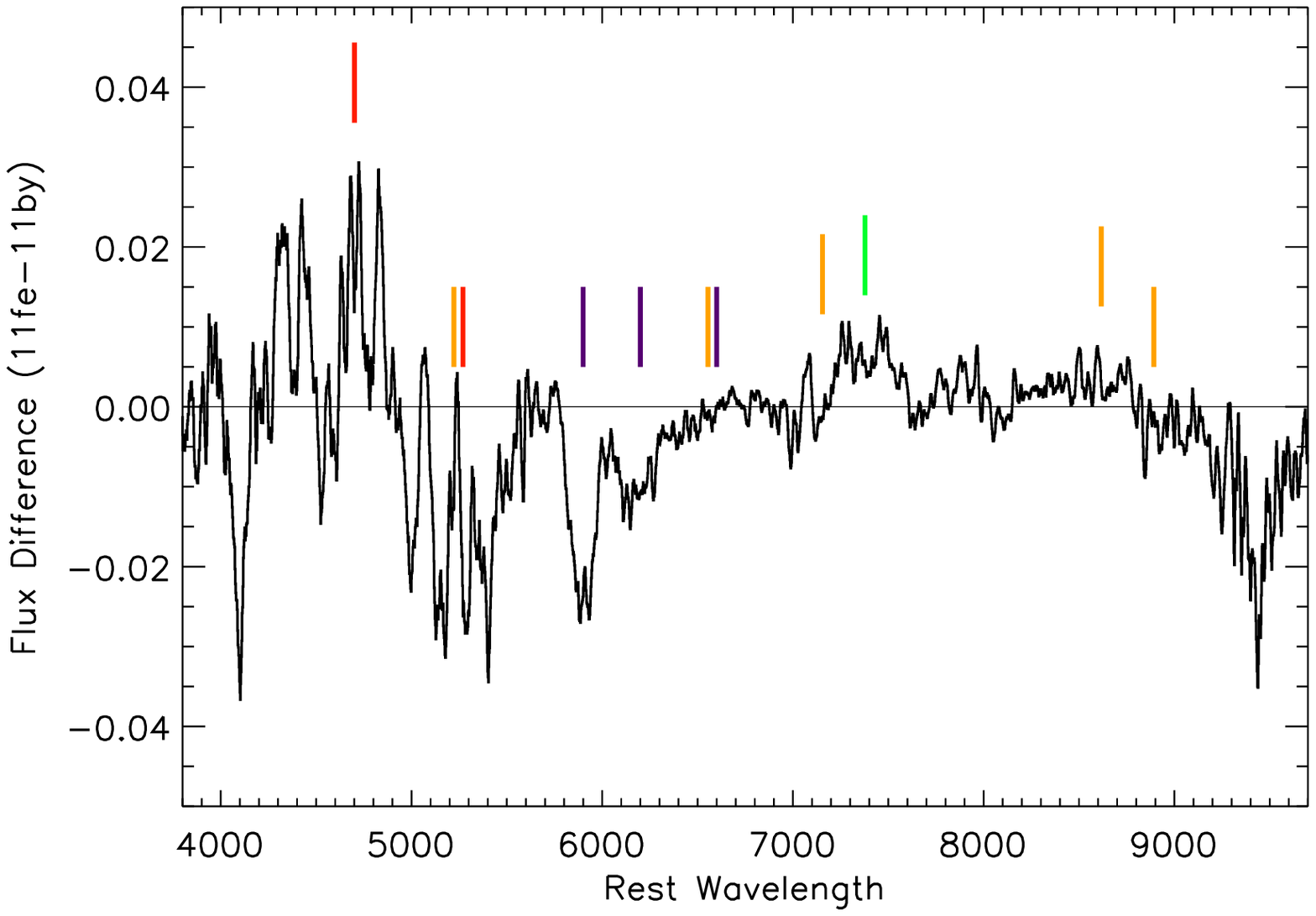}
\includegraphics[trim=0cm 0cm 0cm 0.84cm,clip=true,width=8.5cm]{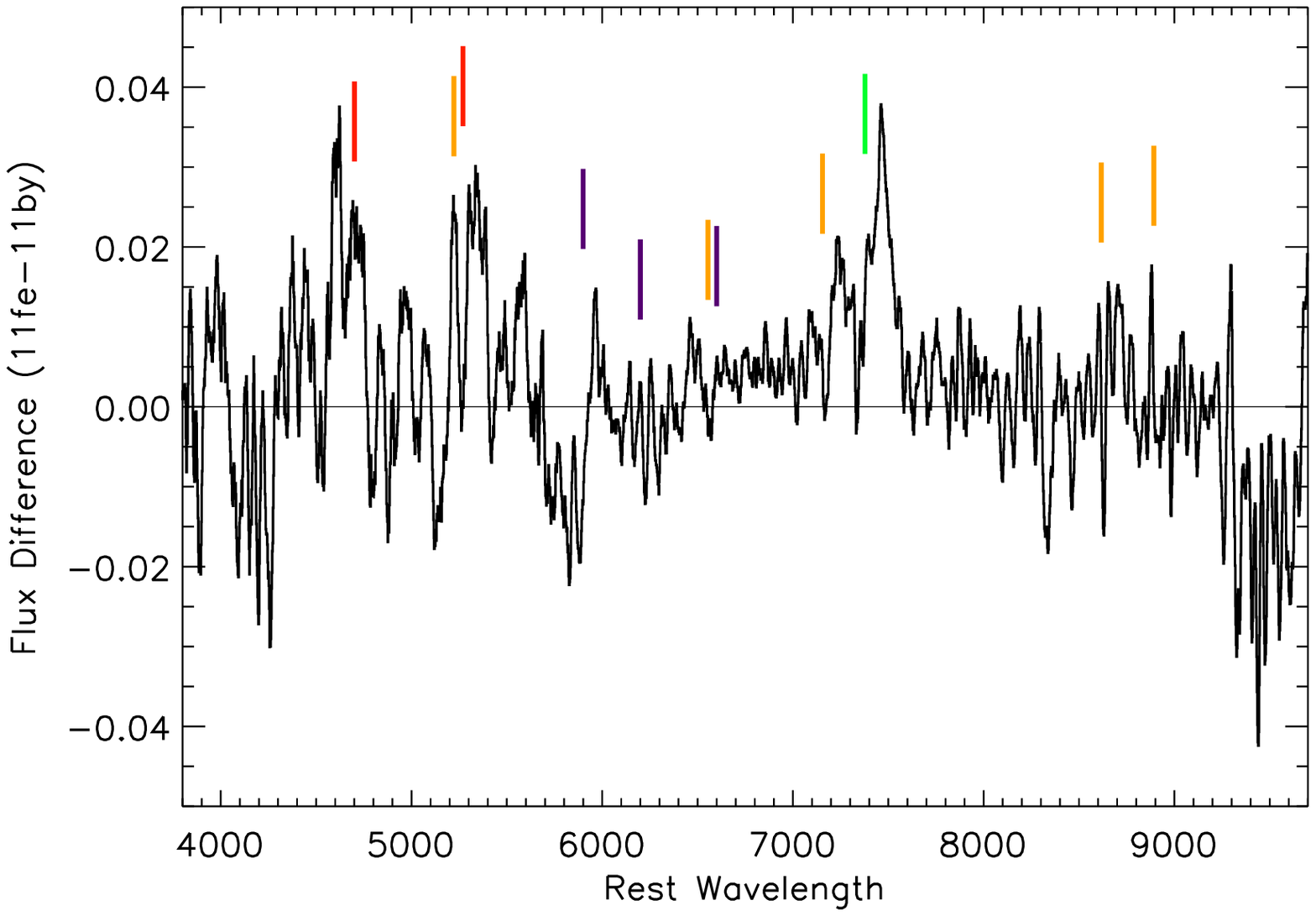}
\caption{Top panels show the late-time nebular phase spectra ($f_\lambda$) of SN\,2011fe (black) and SN\,2011by (blue) at +200 (left) and +300 (right) days past their respective peak brightnesses. These data have been normalised to have equal integrated flux between 4000 and 9500 \AA, and smoothed with a least-squares polynomial smoothing filter of width 20 \AA. The dominant species of forbidden lines are identified with coloured bars. Bottom panels show the difference of the flux-normalised spectra in the top panel (i.e., $f_{\rm fe}-f_{\rm by}$).  \label{fig:nebspec0}}
\end{figure*}

We have established that the nebular spectra of SNe\,2011fe and 2011by are more similar to each other than to other SNe\,Ia. Now we look for subtle distinctions that may reveal differences in their explosions and/or nucleosynthetic products. In the top panels of Figure \ref{fig:nebspec0} we show the full nebular spectra of SN\,2011fe (black) and SN\,2011by (blue), and in the bottom panels their residual spectra, $f_{\rm fe}-f_{\rm by}$. These spectra are dereddened for Galactic extinction along their respective lines of sight (but no host-galaxy extinction, as justified in \S \ref{sec:obs}). Analysis of a residual spectrum is sensitive to errors in the wavelength calibrations and/or a velocity offset between the SNe and their host-galaxy recession velocity. 

We must also consider that all nebular spectra were taken on instruments that acquire the blue and red sides individually (the join is at $\sim 5500$ \AA), and then combined post-reduction. By running the reduction and blue-red join with a variety of parameters, we find that a $\sim5\%$ uncertainty in the relative flux scaling of the blue and red sides is introduced by a choice of overlap region and standard star. 

To minimise the influence of these potential systematics on our residual spectrum, we choose the recession velocity, and the blue and red flux scale parameters, for SN\,2011by that minimise the sum of the squared residuals ($\sum{ ( f_{\rm fe}-f_{\rm by})^2}$). Although our adopted recession velocities for SN\,2011by differs by 200 $\rm km\ s^{-1}$ between the +200 and +300 day spectra, this is only 5 \AA\ at 6000 \AA\ (barely perceptible in the resulting plots). We also find that the difference in the blue and red flux scalings is small, just 1--5\%.

In Figure \ref{fig:nebspec0}, the dominant species contribution to each major line is identified with coloured bars. Despite the general nebular phase similarity, these residual spectra reveal three subtle distinctions. The first is in the [\ion{Fe}{III}] line at $\lambda\approx 4700$ \AA: at both +200 and +300 days, SN\,2011fe exhibits slightly ($\sim5\%$) more flux in this line than does SN\,2011by. The second is in the [\ion{Co}{III}] lines at $\lambda\approx 6000$ \AA: at +200 days, SN\,2011fe exhibits relatively less flux in this line than does SN\,2011by, but this difference is almost gone by +300 days. The third is the [\ion{Ni}{II}] line at $\lambda\approx 7400$ \AA: at +200 days, SN\,2011fe exhibits relatively more flux in this line than does SN\,2011by, and this difference appears even larger at +300 days. To test the robustness of these distinctions, we rerun our residual minimisation routine with these regions excluded. For the first distinction at [\ion{Fe}{III}], the small flux excess exhibited by SN\,2011fe is larger at +200 days but nonexistent at +300 days, so we cannot conclude there is any measurable difference in the [\ion{Fe}{III}] line. For the second distinction at [\ion{Co}{III}], the flux excess exhibited by SN\,2011by is unchanged, and therefore likely real. For the third distinction at [\ion{Ni}{II}], we find that the +300 day flux excess of SN\,2011fe is slightly (2--4\%) larger, and also likely real. We therefore conservatively consider the flux excess of SN\,2011by at +200 days in [\ion{Co}{III}], and the flux excess of SN\,2011fe at +300 days in [\ion{Ni}{II}], to be true intrinsic differences in the nebular phase spectra.

In the scenario where SN\,2011fe synthesised $1.7_{-0.5}^{+0.7}$ times as much radioactive $^{56}$Ni as SN\,2011by, and under the two assumptions that (1) these lines respectively represent the radioactive and stable products and (2) there is a fixed mass budget for SNe\,Ia, we expect SN\,2011fe to exhibit stronger [\ion{Co}{III}] and weaker [\ion{Ni}{II}] lines than SN\,2011by. In both cases we observe the opposite, and since these lines are both on the red-side spectrum, this cannot be explained by the $\sim5\%$ uncertainty in the relative flux scaling discussed above. The observed differences in the nebular residual spectra that we have established here are interpreted with physical models in \S \ref{ssec:discneb}.

\begin{table*}
\begin{center}
\begin{minipage}[bp]{5.2in}
\setlength{\tabcolsep}{2.pt}
\caption{Nebular Spectra Parameters \label{tab:nebspecpars}}
\centering
  \begin{tabular}[bp]{@{}llcccc@{}} 
  \hline
  \hline
Parameter & Line & SN\,2011fe & SN\,2011by & SN\,2011fe & SN\,2011by \\
 & & +224 days & +206 days & +309 days & +310 days \\
\hline
Velocity [km s$^{-1}$] & [FeIII]~$\lambda$4701 & -1360 $\pm$ 91 & -1500 $\pm$ 74 & -774 $\pm$ 64 & -533 $\pm$ 56  \\
($\pm$3$\sigma$)       & [FeII]~$\lambda$7155  & -1190 $\pm$ 180 & -1060 $\pm$ 160 & -1310 $\pm$ 95 & -936 $\pm$ 92  \\
                       & [CoIII] $\lambda$5891 & -609 $\pm$ 120 & -469 $\pm$ 110 & \ldots & -262 $\pm$ 280  \\
                       & [NiII] $\lambda$7378  & -819 $\pm$ 290 & -711 $\pm$ 230 & -638 $\pm$ 140 & -567 $\pm$ 120  \\
\\
FWHM [\AA]       & [FeIII]~$\lambda$4701  & 177 $\pm$ 3 & 180 $\pm$ 3 & 173 $\pm$ 2 & 170 $\pm$ 2  \\
($\pm$3$\sigma$) & [FeII]~$\lambda$7155   & 188 $\pm$ 10 & 194 $\pm$ 10 & 166 $\pm$ 5 & 172 $\pm$ 5  \\
                 & [CoIII]~$\lambda$5891  & 182 $\pm$ 5 & 172 $\pm$ 5 & \ldots & 210 $\pm$ 14  \\
                 & [NiII]~$\lambda$7378   & 211 $\pm$ 15 & 183 $\pm$ 13 & 235 $\pm$ 9 & 193 $\pm$ 8  \\
\\
Flux Ratio       & [NiII]/[FeII]  & 0.77 $\pm$ 0.14 & 0.60 $\pm$ 0.11 & 1.19 $\pm$ 0.12 & 0.89 $\pm$ 0.10  \\
($\pm$3$\sigma$) & [NiII]/[CoIII] & 1.06 $\pm$ 0.16 & 0.78 $\pm$ 0.12 & 3.69 $\pm$ 0.63 & 3.01 $\pm$ 0.55  \\
                 & [CoIII]/[FeII] & 0.73 $\pm$ 0.09 & 0.78 $\pm$ 0.10 & 0.32 $\pm$ 0.05 & 0.30 $\pm$ 0.05 \\
\hline
\end{tabular}
\end{minipage} \end{center}
\end{table*}

\subsubsection{Emission-Line Parameter Fits}\label{sssec:nebeval}

We have shown that SNe\,2011fe and 2011by exhibit twin-like nebular spectra which do not support a $^{56}$Ni mass ratio of $1.7_{-0.5}^{+0.7}$, and that the largest residuals occur at the [\ion{Co}{III}] and [\ion{Ni}{II}] lines. In order to interpret these subtle deviations as physical differences in the explosions of SNe\,2011fe and 2011by, we fit several of the strongest, most distinct lines with Gaussian profiles to estimate the line parameters of width, velocity, and integrated flux.

The emission lines of [\ion{Fe}{III}] $\lambda$4701 and [\ion{Co}{III}] $\lambda$5891 are fit with single Gaussians, while the overlapping lines of [\ion{Fe}{II}] $\lambda$7155 and [\ion{Ni}{II}] $\lambda$7378 are simultaneously fit with a double Gaussian. These lines were chosen because they are relatively unblended (or can be deblended), and represent both stable and radioactive decay products of the rapid nucleosynthesis that occurred in the explosion. Prior to fitting, we correct the spectra for Galactic extinction and redshift. We use the recession velocity for SN\,2011by that best minimises the nebular spectra residuals at both +200 and +300 days: $v_{\rm rec}=500\,\rm km\ s^{-1}$. This is a difference of $\sim350\,\rm km\ s^{-1}$ from the measured recession velocity of the host of SN\,2011by, $v_{\rm rec}=852\,\rm km\ s^{-1}$.

In order to minimise the contamination from neighbouring lines we subtract a pseudo-continuum created by joining the local minima on each side of the emission line. This is suitable for our purposes because we do not calculate intrinsic line luminosities or derive physical parameters, but only consider evolution in relative line flux ratios. We perform a least-squares fit to the non-normalised, flux-calibrated spectra and bootstrap our parameter uncertainties. In Table \ref{tab:nebspecpars} we list the Gaussian parameters and their $3\sigma$ uncertainties. The [\ion{Co}{III}] $\lambda$5891 line in the +309 day spectrum of SN\,2011fe is asymmetric and not well represented by a Gaussian, so we do not use the fit line width or velocity in our analysis.

\textbf{Velocity ---} The emission-line velocity is calculated from the fit peak wavelength ($\lambda_{\rm p}$) of the Gaussian using the relativistic Doppler equation, where the rest wavelengths of each of the fit emission features are given in Table \ref{tab:nebspecpars}. Considering the 3$\sigma$ uncertainties in the fit peak wavelengths, and also the potential $\lesssim350\,\rm km\ s^{-1}$ systematic introduced by using the recession velocity that minimises the residuals, all we can say with confidence is that the iron lines are blueshifted with respect to the cobalt and nickel lines. Maeda et al. (2010) show that for an asymmetric SN\,Ia explosion, the observer's viewing angle (i.e., the angle between the line-of-sight vector and the vector from progenitor core to detonation centre) affects the observed \ion{Si}{II} velocity gradient and nebular line velocities. If the detonation centre is on the near/far side of the star with respect to the observer, the SN\,Ia displays \ion{Si}{II} with a low/high velocity gradient in the photospheric spectra and blueshifted/redshifted iron lines in the nebular spectra. The blueshift in nebular iron lines and the low \ion{Si}{II} velocity gradient that we observe for both SNe\,2011fe and 2011by suggest that if they are asymmetric explosions, we are viewing them from approximately the same angle --- and we note that there is evidence for asymmetry from spectropolarimetry of SN\,2011fe (Smith et al. 2011). We also note that the temporal evolution of the [\ion{Fe}{III}] $\lambda$4701 line toward zero velocity, as pointed out by Silverman et al. (2013), is displayed by both SNe\,2011fe and 2011by.

\textbf{FWHM ---} The line widths are represented as the full width at half-maximum intensity (FWHM) of the fit Gaussian, and quoted with 3$\sigma$ uncertainties in the fit value. The FWHM of the iron emission lines in SNe\,2011fe and 2011by are equivalent within 3$\sigma$ in both the +200 and +300 day epochs, but the cobalt and nickel lines are wider in SN\,2011fe at $>3\sigma$ significance. This suggests that SN\,2011fe formed both radioactive and stable nickel in larger volumes. This agrees with the results of Mazzali et al. (2014), who model the early-time NUV-optical spectral evolution of SN\,2011fe and find that an extended region of $^{56}$Ni is required for SN\,2011fe. For both SNe\,Ia, between +200 and +300 days the iron lines appear to become narrower, and the cobalt and nickel lines become wider --- but we caution that this may simply be due to the difference in the psuedo-continuum, and the changing relative contamination from neighbouring lines at these two epochs. 

\textbf{Flux Ratios ---} We select three lines, all from the red side of the spectrum, to compare with line flux ratios: [\ion{Ni}{II}] (representing stable nickel), [\ion{Co}{III}] (representing radioactive nickel), and [\ion{Fe}{II}] (representing a combination of stable iron and radioactive nickel). Owing to the uncertainty in the relative flux scaling between the blue and red sides, we do not compare features on opposite sides of the chip gap at $\lambda\approx 5500$ \AA\ (i.e., we exclude [\ion{Fe}{III}]). We find that the [\ion{Ni}{II}]/[\ion{Fe}{II}] ratio is consistent between SNe\,2011fe and 2011by at +200 days, but larger for SN\,2011fe at +300 days (at $>3\sigma$). In contrast, the [\ion{Ni}{II}]/[\ion{Co}{III}] ratio is larger for SN\,2011fe at +200 days (at $>3\sigma$) and is consistent at +300 days, but at this time the uncertainties are much larger because of the decline in flux of the cobalt line. Finally, we find that the [\ion{Co}{III}]/[\ion{Fe}{II}] ratio is consistent at both +200 and +300 days. This seems to contradict Figure \ref{fig:nebspec0}, where the [\ion{Co}{III}] line for SN\,2011by at +200 days exhibits an excess in peak flux over SN\,2011fe. However, we do not see an excess in integrated [\ion{Co}{III}] line flux because the line is narrower in SN\,2011by than in SN\,2011fe. 

At nebular epochs the line flux is governed by both the quantity and ionisation temperature of the emitting material. As previously discussed, if SN\,2011fe synthesised a $^{56}$Ni mass $1.7_{-0.5}^{+0.7}$ times greater than SN\,2011by, then it should have more flux in [\ion{Co}{III}] (radioactive product) and less flux in [\ion{Ni}{II}] (stable product) --- but we observe the opposite. Our physical interpretations of these observations, including a discussion of the role of ionisation temperature in the nebula, are reserved for discussion in \S \ref{ssec:discneb}.

\section{Discussion} \label{sec:disc}

In this section we utilise theoretical and analytical models of SNe\,Ia to determine the most accurate physical interpretation of the similarities and differences exhibited by SNe\,2011fe and 2011by. To begin, we consider FK13's comparison of near-maximum NUV-optical spectra to the models of Lentz et al. (2000), which represent a range of progenitor metallicities. They found that the lower NUV flux level for SN\,2011by is consistent with it having a higher progenitor metallicity. With this work we extend the comparison to pre-maximum spectra (Figure \ref{fig:specearly}) and find that at 10 days before peak brightness the NUV flux discrepancy is stronger and extends to redder wavelengths. Both of these effects are predicted by the models of Lentz et al. (2000), further supporting a large difference in metal abundance between the progenitor stars of these two otherwise twin-like SNe\,Ia. 

The derived values for the progenitor metallicities are presented and discussed in \S \ref{ssec:discphot}, including the revisions required if the Tully-Fisher distance modulus for SN\,2011by is an underestimate. In \S \ref{ssec:discneb} we demonstrate that the nebular phase spectra do not support a difference of $\sim 0.2\, \rm M_{\odot}$ in $^{56}$Ni mass, but instead reinforce the prospect that the Tully-Fisher distance is incorrect, and that SNe\,2011fe and 2011by reached similar peak absolute magnitudes. Section \ref{ssec:discaltpm} broadens the support for progenitor metallicity by comparing with more recent spectral models than Lentz et al. (2000), and in \S \ref{ssec:discalt} we strengthen it by considering (but ultimately rejecting) a wider variety of alternative physical causes for the NUV flux depression. We discuss in \S \ref{ssec:discLC} the possible physical causes for the slower decline leading to a late-time luminosity excess for SN\,2011by.

\subsection{The Progenitor Metallicity Interpretation \\ for SN\,2011fe and 2011by}\label{ssec:discphot}

As previously discussed by FK13 and in \S \ref{ssec:anaphot}, at peak brightness SN\,2011by appeared intrinsically fainter than SN\,2011fe by $\sim0.6$\,mag in optical bands, indicating that it produced $\sim0.2\,\rm M_{\odot}$ less $^{56}$Ni (assuming the Tully-Fisher distance modulus for SN\,2011by). SN\,2011by also exhibited a lower NUV flux, which is indicative of a higher metal abundance than SN\,2011fe. A higher metallicity progenitor has more neutrons available and can synthesise more stable $^{54}$Fe and $^{58}$Ni, but it produces less radioactive $^{56}$Ni (Timmes et al. 2003). Where $M(\rm{^{56}Ni})$ is the synthesised mass of $^{56}Ni$, and $Z$/$\rm Z_{\odot}$ is the progenitor star's initial metallicity while it was on the main sequence, Timmes et al. (2003) show that

\begin{equation}
M(^{56}{\rm Ni}) \propto 1 - 0.057\ Z / {\rm Z}_{\odot}.
\end{equation}

\noindent
By this relation, the $\sim0.2\, \rm M_{\odot}$ difference in $^{56}$Ni implies progenitor metallicities of $Z_{\rm by}$/$\rm Z_{\odot}$ = $Z_{\rm fe}$/$\rm Z_{\odot}$ + 3.5. For SN\,2011fe, a subsolar metallicity is implied by its host galaxy (\S \ref{ssec:host}), and supported by Mazzali et al. (2014), who apply abundance tomography models to the NUV-optical $HST$ spectra and find that the amount of iron in the outer layers of SN\,2011fe is consistent with subsolar metallicity, $Z_{\rm fe}\approx 0.5$ $\rm Z_{\odot}$. FK13 show that together, this implies a progenitor metallicity for SN\,2011by of $Z_{\rm by}\approx 4\,\rm Z_{\odot}$.

However, if the Tully-Fisher distance to NGC 3972 underestimates the true distance to SN\,2011by, then both SNe\,Ia synthesised comparable masses of $^{56}$Ni. How do we reconcile this with the lower NUV flux of SN\,2011by, which corresponds to a factor of $\sim 30$ higher number abundance of elements heavier than oxygen (e.g., the Lentz model represented by the orange line in Figure \ref{fig:specearly})? This combination of a small difference in $M$($^{56}$Ni) and a large difference in the number abundance of metals would indicate that the metallicities of both progenitors were significantly subsolar\footnote{The assumption here is that the metallicity of the main-sequence progenitor star propagates through all stages of stellar evolution, as the hydrogen is burned to the carbon and oxygen, to a comparable abundance of heavy metals in the white dwarf's outer layers.}. For example, if $Z_{\rm fe} = 0.03\, \rm Z_{\odot}$ and $Z_{\rm by} = 0.90\, \rm Z_{\odot}$, then $M_{\rm fe}(^{56}{\rm Ni}) - M_{\rm by}(^{56}{\rm Ni})$ = 0.05\, $\rm M_{\odot}$, which corresponds to a difference in absolute peak magnitude of $\sim0.1$. This kind of constraint on the metallicities will only be possible when we verify the distance to the host of SN\,2011by, NGC 3972, using Cepheid variables with $HST$ in Cycle 22.

\subsection{Nebular Spectra and the \\ Progenitor Metallicity Interpretation} \label{ssec:discneb}

If SN\,2011by synthesised $\sim0.2\, \rm M_{\odot}$ less $^{56}$Ni than SN\,2011fe owing to its higher progenitor metallicity, then under the assumption of a fixed total mass SN\,2011by instead formed $\sim0.2\,\rm M_{\odot}$ of more stable products. In this scenario, the ratio of $^{54}$Fe to $^{58}$Ni is $\sim2$ (Timmes et al. 2003), which implies that SN\,2011by formed $\sim0.13\, \rm M_{\odot}$ of stable $^{54}$Fe and $\sim0.07\, \rm M_{\odot}$ of $^{58}$Ni. This is in addition to the base amount of stable $^{54}$Fe and $^{58}$Ni made in both SNe, $\sim0.2\,\rm M_{\odot}$ in each given that they are both ``normal" SNe\,Ia (Mazzali et al. 2007). The iron in nebular phase spectra is a combination of stable $^{54}$Fe and the product of the radioactive decay of $^{56}$Ni, whereas the nickel is representative of the amount of stable $^{58}$Ni. Under this ``toy model" accounting, SN\,2011by should have $\sim2$ times more stable nickel than SN\,2011fe, and a similar amount of iron. We should see evidence of stronger [\ion{Ni}{II}], and stronger nickel-to-iron and nickel-to-cobalt line flux ratios, in SN\,2011by compared to SN\,2011fe.

In contrast to this expectation, the nebular phase spectra are remarkably similar, and SN\,2011by actually demonstrates weaker [\ion{Ni}{II}], and lower nickel-to-iron and nickel-to-cobalt line flux ratios. The nebular phase spectra of SNe\,2011fe and 2011by do not support this ``toy model" in which progenitor metallicity strongly influences the nucleosynthetic products, but there are three important points to consider here. (1) Assuming that the progenitor metallicity caused the entire $\sim0.2\, \rm M_{\odot}$ difference in $^{56}$Ni extends beyond the $\sim10\%$ variation in $^{56}$Ni mass typically associated with progenitor metallicity (Kasen et al. 2009). (2) Our line identification may be too facile, as it does not consider blending with lines from other elements such as titanium and calcium, the contributions from which remain an open question. (3) The ionisation temperature of the nebula also plays a role in strength of nebular lines. For example, McClelland et al. (2013) examine the late-time decline of SN\,2011fe and find that the line flux in doubly ionised species declines more rapidly than that of singly ionised species, because the decreasing temperature of the nebula leaves less energy available for ionisation. As discussed in \S \ref{sssec:late}, the photometric decline of SN\,2011by is slower than that of SN\,2011fe, leading to an excess brightness by +200 days past maximum. This may indicate that the temperature of SN\,2011by is declining more slowly than that of SN\,2011fe, which could happen if the iron-rich material in SN\,2011by is more efficient at trapping the positrons and/or gamma-rays from the $\beta$-decay of $^{56}$Co. 

We find that the nebular spectra suggest the nucleosynthetic yields are quite similar for SNe\,2011fe and 2011by, but we cannot rule out a contrivance of different $^{56}$Ni masses with line blending and ionisation temperatures to create such twin-like nebular spectra. It is important to note that the scenario in which SN\,2011fe and 2011by experienced similar physical explosions is {\it not} inconsistent with a relatively large difference in their progenitor metallicities, as it may seem to be. Additional line blanketing by enriched material in the outer layers of SN\,2011by still occurs, but in this case both SNe have low metallicity, and the imbalance in neutronisation rates during the explosions are not so different that the nucleosynthetic yields are affected.

\subsection{Alternative Progenitor Metallicity Models for the NUV-Optical Photospheric Spectra}\label{ssec:discaltpm}

The NUV flux depression of SN\,2011by relative to SN\,2011fe has hitherto only been interpreted with the progenitor metallicity models of Lentz et al. (2000). We now extend this discussion to include the more recent models of Sauer et al. (2008) and Walker et al. (2012). Sauer et al. (2008) apply radiative transfer models to synthesise spectra for SNe\,Ia 2001eh and 2001ep, which also exhibit a NUV flux difference but are otherwise less twin-like than SNe\,2011fe and 2011by. They find that a higher iron content in the progenitor atmosphere can produce more NUV flux owing to the scattering of photons from red to blue wavelengths --- the opposite of Lentz et al. (2000). Under this interpretation, SN\,2011fe had the higher metallicity progenitor. However, Sauer et al. (2008) also find that the photon scattering leads to a depressed flux at wavelengths $4000\lesssim\lambda\lesssim5000$ for higher metallicity progenitors, and also to large variations at $\sim3500$ \AA, that are not seen in SNe\,2011fe and 2011by. 

\begin{figure}
\begin{center}
\includegraphics[width=8.5cm]{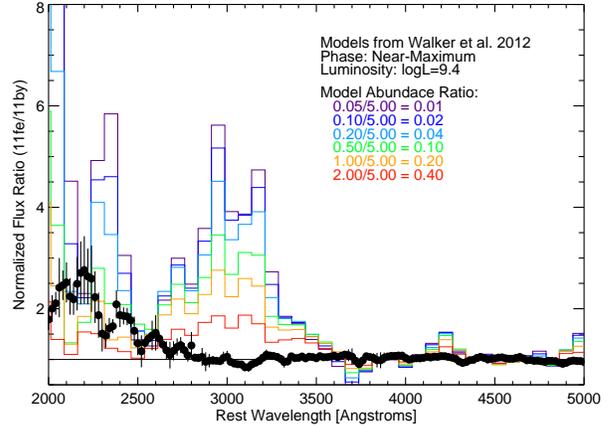}
\caption{The flux-ratio spectrum: SN\,2011fe divided by SN\,2011by after being normalised between 4000 and 5500 \AA, and binned by 20 \AA\ (black circles; error bars represent standard deviation in the bin). Coloured histograms represent the flux ratio of near-maximum model spectra with different abundances from Walker et al. (2012). The flux ratio is the lower divided by the higher metallicity progenitor model; colours from purple to red represent a decreasing metallicity difference. Compared with the lower panel of Figure \ref{fig:specearly}, the Walker et al. (2012) models do not match the observed NUV flux discrepancy as well as those from Lentz et al. (2000). \label{fig:walker}}
\end{center}
\end{figure}

Walker et al. (2012) use observations of the normal SN\,Ia 2005cf, which had $\Delta m_{15}(B)=1.10$\,mag, to create model spectra for progenitors with a variety of metallicities. In Figure \ref{fig:walker} we compare the flux-ratio spectrum of SN\,2011fe to 2011by with flux-ratio spectra from models with bolometric luminosity $\log{L/{\rm L}_{\odot}}=9.4$. We use the highest abundance model\footnote{$\eta=5.0$, where $\eta$ is the multiplicative factor for the abundance of elements heavier than calcium; $\eta =$ 0.05, 0.1, 0.2, 0.5, 1, 2, and 5. A given metallicity ratio, such as $\eta_1/\eta_2 = 0.1$, can be obtained with (for example) 0.05/0.5, 0.1/1.0, 0.2/2.0, or 0.5/5.0 --- all flux-ratio spectra are similar, and only a relative abundance can be inferred.} as the denominator in order to assess the maximum possible effect. We find that increasing the progenitor metallicity leads to a larger NUV flux depression, which is broadly consistent with the Lentz et al. (2000) models. Walker et al. (2012) predict a greater difference around 3000 \AA\ in near-maximum spectra than observed (although we do see a smaller feature at this location in the pre-max spectra; e.g., Figure \ref{fig:kasen_mu}), but the overall relation between NUV flux and progenitor metallicity is supported.

The Lentz, Sauer, and Walker models also predict that progenitor metallicity causes small but distinguishable variations in the silicon and carbon lines in the photospheric optical spectra. First, the \ion{Si}{II} $\lambda$6355 line velocity appears to, in turn, be either blueshifted in higher metallicity progenitors (Lentz et al. 2000), relatively unchanged (Sauer et al. 2008), or redshifted in progressively higher metallicity progenitors (Walker et al. 2012). As established in \S \ref{ssec:anaspec}, the \ion{Si}{II} $\lambda$6355 line evolution is very similar in SNe\,2011fe and 2011by. Despite being a notable feature of the models, this apparent relation between progenitor metallicity and silicon line velocity is probably model dependent and not a constraining feature. Second, the higher metallicity progenitor that creates less $^{56}$Ni may also exhibit a greater amount of unburned carbon. We find that both SNe\,2011fe and 2011by show carbon at similar phases, which suggests they have similarly sized regions of unburned carbon.

To summarise, models of SN\,Ia progenitors with varying atmospheric metal abundances do not agree on the size or sign of the effect on NUV flux, only that it will be affected. They furthermore do not agree on potential changes in the optical photospheric absorption lines, but do agree that they are less significant than the effect on NUV flux. In this way, the models do support progenitor metallicity as a plausible explanation for the NUV discrepancy between the optically twin SNe\,Ia 2011fe and SN\,2011by.

\subsection{Alternative Physical Interpretations for \\ the NUV Flux Discrepancy} \label{ssec:discalt}

Here we consider alternate physical mechanisms that could produce the NUV flux discrepancy  between SNe\,2011fe and 2011by that has thus far been attributed to metallicity. This part of the discussion is agnostic to the disputed distance modulus and $^{56}$Ni mass of SN\,2011by.

\textbf{Distribution of Nucleosynthetic Products ---} If nucleosynthetic burning occurs in the outer layers of the white dwarf, the newly synthesised metals could create extra opacity and lead to an NUV flux deficit. This may happen if there is a preliminary detonation in the surface layer of accreted material, known as the double-detonation scenario (Woosley \& Weaver 1994; Fink et al. 2007), or if the nucleosynthetic processes happened relatively near the surface (Piro 2012). However, both cases likely cause significant differences in the physical explosion that would affect the optical luminosity and spectra, such as a steeper light-curve rise (Piro 2012) and high-velocity burned material (Hachinger et al. 2013). No difference in these qualities is observed for SNe\,2011fe and 2011by.

\textbf{Kinetic Energy from $^{56}$Ni Mass ---} An explosion with a higher mass of synthesised $^{56}$Ni could impart a larger kinetic energy to its ejecta, shifting more metal lines into the NUV and depressing the NUV flux. This effect is seen in the models of Sauer et al. (2008), and Maguire et al. (2012) show that brighter SNe\,Ia exhibit blueshifted NUV spectral lines and a depressed NUV flux with respect to fainter SNe\,Ia. However, a kinetic-energy difference would influence the velocity evolution of the \ion{Si}{II} $\lambda$6355 line, which we observe to be the same in both of our SNe\,Ia.

\textbf{Ejected Mass ---} If the total mass of a white dwarf prior to explosion is not fixed at the Chandrasekhar mass, then an explosion with a lower total mass may impart a {\it relatively} larger amount of kinetic energy to its ejecta, blueshift more metal lines, and exhibit a depressed NUV flux. However, this scenario would also affect the velocity of other photospheric absorption lines where we see good agreement between SNe\,2011fe and 2011by.

\begin{figure}
\includegraphics[width=8.5cm]{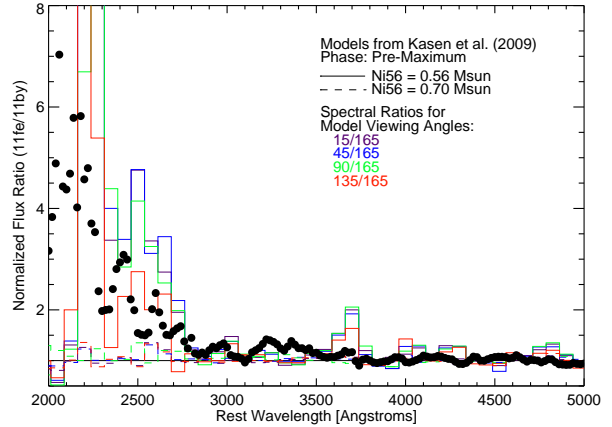}
\caption{The flux ratio spectrum: SN\,2011fe divided by SN\,2011by, after being normalised between 4000 and 5500 \AA, and binned by 20 \AA\ (black circles; error bars represent standard deviation in the bin). Coloured lines represent the spectral flux of a model with viewing angle $15^circ$, $45^\circ$, $90^\circ$, or $135^\circ$ (colours), divided by one with a viewing angle of $165^\circ$, where both models have the same mass of $^{56}$Ni (Kasen et al. 2009). These models have been normalised in the optical, are for $-9$ days pre-maximum, and are shown for two $^{56}$Ni masses (solid and dashed lines). \label{fig:kasen_mu}}
\end{figure}

\textbf{Viewing Angle ---} For the case of an isotropic distribution of ignition points (i.e., a symmetric explosion), Blondin et al. (2011) show that synthesised spectra from Kasen et al. (2009) display discrepancies at $<3500$ \AA\ that are correlated with viewing angle, while remaining similar at optical wavelengths. In Figure \ref{fig:kasen_mu}, we compare the flux-ratio spectrum of SNe\,2011fe and 2011by to model spectra for a variety of viewing angles. The model spectra are flux normalised at optical wavelengths, and divided by the one with the lowest NUV flux (viewing angle $165^\circ$). We have used two values of $^{56}$Ni mass in order to show that this effect is stronger for SNe\,Ia that synthesise less nickel. From Figure \ref{fig:kasen_mu} it seems that viewing angle is a promising explanation for the NUV flux discrepancy, but Blondin et al. (2011) state that the $U$-band flux excess of their models is actually a problem with flux redistribution into the infrared, and furthermore point out that their models produce photospheric phase spectral features with high velocity.

\textbf{Outer Layer Density Gradient ---} Sauer et al. (2008) show that varying the progenitor's outer layer density gradient affects the NUV flux: a shallower density gradient leads to higher densities in the outer layers, which increases the line opacity and causes more efficient line blocking at NUV wavelengths. The result is a depressed NUV flux at wavelengths less than $\sim3000$ \AA, with little effect on the optical flux. In contrast, several metallicity models predict variations at $\sim3000$ \AA\ that are not seen in SNe\,2011fe and 2011by. Under this interpretation, SN\,2011by would have a shallower density gradient, and denser material in its outer layers. This may result from a longer deflagration before transition to detonation in SN\,2011by, which could also produce the narrower nebular phase emission lines discussed in \S \ref{ssec:discneb}. A lower density for SN\,2011fe agrees with Mazzali et al. (2014), whose custom physical model requires a lower density than that of a typical delayed detonation. However, different outer layer density gradients for SNe\,2011fe and 2011by would also be expected to change the \ion{Si}{II} velocity, which we find agrees very well between SNe\,2011fe and 2011by (e.g., Figure \ref{fig:vel_Si6150}).

\textbf{Summary ---} We find that most alternatives to progenitor metallicity that affect the NUV flux of a SN\,Ia are also expected to have an influence on the photometry or optical spectra. Given the near twinness of SNe\,2011fe and 2011by in these respects, progenitor metallicity remains the simplest explanation for the NUV flux discrepancy, but there is no way to rule out partial contributions from these alternatives.

\subsection{Physical Constraints from the Photometric Evolution of SN2011by}\label{ssec:discLC}

In \S \ref{sssec:late} we established that SN\,2011by exhibited a slower decline than SN\,2011fe, which led to a luminosity excess by +200 days past maximum that could not be explained by a light echo, CSM interaction, or host-galaxy contamination. Here we discuss the physical interpretation of this slower decline with respect to the true distance to SN\,2011by.

\noindent
\textbf{If the Tully-Fisher distance is accurate ---} In this scenario, the absolute peak magnitude of SN\,2011by was $\sim0.6$\,mag fainter than SN\,2011fe, but equivalent at late times. SN\,2011by may be more efficient at trapping the energy from cobalt decay and reradiating it as optical luminosity, and/or a significant amount of energy at late times could also come from the longer-lived isotopes $^{57}$Co (272 days) or $^{44}$Ti (59 years; Seitenzahl et al. 2009). Alternatively, the equivalent late-time luminosity may indicate that SNe\,2011fe and 2011by formed a similar amount of $^{56}$Ni, but that the peak brightness of SN\,2011by was depressed. A larger ejecta mass for SN\,2011by could lower the peak brightness, but would also increase the diffusion time (rise time), and change the shape of the light curve (Arnett 1982). Ultimately we find such a violation of Arnett's rule difficult to physically explain, especially given the similar optical spectra. If the Tully-Fisher distance modulus is correct, the most likely explanation is that SN\,2011fe formed more $^{56}$Ni, and that SN\,2011by has more efficient trapping of energy at late times.

\noindent
\textbf{If the Tully-Fisher distance is an underestimate ---} In this scenario, SNe\,2011fe and 2011by reached similar absolute peak magnitudes and produced more equivalent amounts of $^{56}$Ni. This would explain their twin-like optical and nebular spectra and light-curve widths. The higher metal abundance for SN\,2011by remains a valid interpretation of its lower NUV flux, but in this case both SNe have low metallicity. In this scenario SN\,2011by is intrinsically brighter than SN\,2011fe at late times, and as before this is likely due to more efficient energy trapping and/or contributions from longer lived isotopes.

\section{Conclusion} \label{sec:con}

Our extended investigation of SNe\,Ia 2011fe and 2011by has led to significantly different conclusions than previous work. We have shown that the remarkably twin-like nebular spectra are best explained by explosions with nearly equivalent nucleosynthetic yields. This implies that SN\,2011fe and 2011by reached similar absolute magnitudes, and we propose that the Tully-Fisher distance modulus for NGC 3972, the host galaxy of SN\,2011by, has been underestimated. To resolve this issue we have been awarded $HST$ time in Cycle 22 to obtain a Cepheid distance modulus.

We have demonstrated that the NUV flux of SN\,2011by is lower than that of SN\,2011fe at epochs $-10$ days before maximum light, and that the progenitor metallicity models of Lentz et al. (2000) support the interpretation of this as a higher metal abundance in SN\,2011by. We extend this analysis to include additional progenitor metallicity models --- they agree that progenitor metallicity will affect the NUV flux, but not on the size or sign of the effect. We also consider alternative sources for NUV flux diversity: the distribution of nucleosynthetic products, kinetic energy, ejected mass, viewing angle, and outer layer density gradient. Although they all predict larger changes in the optical spectra (e.g., in the \ion{Si}{II} velocity gradient) than progenitor metallicity, they may contribute. We have several reasons to suspect that density plays a role, such as narrower nebular lines and a slower late-time decline in SN\,2011by, and the lower density and larger volume for the nucleosynthetic products of SN\,2011fe demonstrated by Mazzali et al. (2014). If SNe\,2011fe and 2011by synthesised equivalent amounts of $^{56}$Ni, but the metal abundance of 2011by was significantly enhanced, then {\it both} SNe\,Ia had $\lesssim \rm Z_{\odot}$ metallicity progenitors. This is consistent with our estimated metallicities at the sites of these SNe in their host galaxies. 

Finally, we have shown that SN\,2011by experienced a slower late-time decline than SN\,2011fe, and that this is most likely caused by more efficient energy trapping or the existence of more longer-lived isotopes. Given the extreme similarity of the features in the nebular phase spectra, the former is more likely and is consistent with the possibility that SN\,2011by was a higher density detonation than SN\,2011fe.

This work is the result of community efforts to assemble and share detailed time series of photometry and spectroscopy for nearby SNe\,Ia, and only through the continuation of these cooperative endeavors can we persevere toward a deeper understanding of these cosmologically valuable standard candles.

\section*{Acknowledgements}

Based on observations from the Space Telescope Imaging Spectrograph with the NASA/ESA {\it Hubble Space Telescope}, the Low Resolution Imaging Spectrometer at the Keck-1 telescope, and (at Lick Observatory) the Katzman Automatic Imaging Telescope, the Nickel telescope, and the Kast spectrograph on the Shane telescope. We are grateful to the staffs at the Lick and Keck Observatories for their
assistance. We thank the dedicated Nickel telescope observers: Michael Kandrashoff, Michael Ellison, Peter Blanchard, Byung Yun Choi, Daniel Cohen, Michelle Mason, Andrew Wilkins, Kyle Blanchard, Chadwick Casper, Kiera Fuller, Gary Li, Daniel Krishnan, and Isha Nayak. The W.~M.\ Keck Observatory is operated as a scientific partnership among the California Institute of Technology, the University of California, and NASA; it was made possible by the generous financial support of the W.~M.\ Keck Foundation. This work has also used the Weizmann Interactive Supernova Data Repository (WISEREP) at www.weizmann.ac.il/astrophysics/wiserep.

We thank Dan Kasen for his helpful discussions and for providing model spectra, Emma Walker for her useful correspondence and for providing model spectra, and Daniel Perley for the use of (and assistance with) his Keck LRIS imaging and spectroscopy reduction pipeline\footnote{http://www.astro.caltech.edu/$\sim$dperley/programs/lpipe.html}. J.M.S. is supported by an NSF Astronomy and Astrophysics Postdoctoral Fellowship under award AST-1302771. The supernova research of A.V.F.'s group at U.C. Berkeley is supported by Gary \& Cynthia Bengier, the Richard \& Rhoda Goldman Fund, the Christopher R. Redlich Fund, the TABASGO
Foundation, NSF grant AST--1211916, and NASA/HST grant GO-13286. KAIT and its ongoing operation were made possible by donations from Sun Microsystems, Inc., the Hewlett-Packard Company, AutoScope Corporation, Lick Observatory, the NSF, the University of California, the Sylvia \& Jim Katzman Foundation, and the TABASGO Foundation.  


\bibliographystyle{apj}
\bibliography{apj-jour,myrefs}

\label{lastpage}

\end{document}